\documentclass[prl,twocolumn,10pt,aps,longbibliography]{revtex4-1}
\usepackage{amsmath,amssymb,bm}
\usepackage{graphicx}
\usepackage{epstopdf}
\usepackage{latexsym}
\usepackage{subfigure}
\usepackage{color}
\usepackage{natbib}
\usepackage{hyperref}
\usepackage{braket}
\usepackage{dsfont}
\usepackage[english]{babel}
\usepackage{blindtext}
\hypersetup{
  colorlinks,
  citecolor=magenta,
  linkcolor=blue,
  urlcolor=blue}

\begin{document}

\title{Entanglement entropy from nonequilibrium work}
\author{Jonathan D'Emidio}
\affiliation{Institute of Physics, \'{E}cole Polytechnique F\'{e}d\'{e}rale de Lausanne (EPFL), CH-1015 Lausanne, Switzerland}
\begin{abstract}

The R\'enyi entanglement entropy in quantum many-body systems can be viewed as the difference in free energy between partition functions with different trace topologies. We introduce an external field $\lambda$ that controls the partition function topology, allowing us to define a notion of nonequilibrium work as $\lambda$ is varied smoothly. Nonequilibrium fluctuation theorems of the work provide us with statistically exact estimates of the R\'enyi entanglement entropy.  This framework also naturally leads to the idea of using quench functions with spatially smooth profiles, providing us a way to average over lattice scale features of the entanglement entropy while preserving long distance universal information.  We use these ideas to extract universal information from quantum Monte Carlo simulations of SU(N) spin models in one and two dimensions.  The vast gain in efficiency of this method allows us to access unprecedented system sizes up to 192 x 96 spins for the square lattice Heisenberg antiferromagnet.
\end{abstract}
\maketitle

Entanglement entropy is a quantity of basic importance in the characterization of quantum many-body wavefunctions.  It signals the departure from a simple tensor product of subsystem wavefunctions for a spatial bipartition.  Though in general the entanglement entropy depends only on the boundary size of the bipartition--the ``area law"--and captures local correlations, violations to the area law can contain long distance universal information [\onlinecite{Amico2008:EE, Eisert2010:AreaLaws, Laflorencie2016:Entanglement}]. For instance, one dimensional critical systems show a logarithmic growth of the entanglement entropy with subsystem size where the prefactor depends on the central charge of the underlying conformal field theory [\onlinecite{Holzhey1994:Geometric, Vidal2003:EinQCP, Calabrese2004:EEinQFT}]. Examples of universal features in higher dimensions include logarithmic contributions in symmetry broken phases with gapless Goldstone modes [\onlinecite{Metlitski2011:GoldstoneEE, Castro-Alvaredo2012:EEhighDeg, Song2011:EEHeisen, Ju2012:2Dgapless, Luitz2015:UniversalLog, Laflorencie2015:SpinWaveEE, Rademaker2015:TowerOfStates, Kallin2011:AnomaliesEEHeisen, Kulchytskyy2015:DetectingGoldstone, Luitz2017:QMClinesubsystem}] and a universal negative constant in topological phases [\onlinecite{Kitaev2006:TEE, Levin2006:TEE, Grover2013:EEPortal}].


Low entanglement methods such as the density matrix renomalization group (DMRG) most naturally allow for the computation of entanglement entropy, which has been enormously successful in characterizing critical spin chains and identifying 2D gapped topological phases of matter [\onlinecite{Laflorencie2016:Entanglement}].  However, 2D gapless systems, long-range interactions and 3D systems still pose significant challenges for the DMRG method.  This motivates the development of entanglement entropy methods in quantum Monte Carlo (QMC) that, barring a sign problem, do not face the same difficulties.

In fact, QMC simulations have made remarkable progress in this direction. The ability to interpret the R\'{e}nyi entanglement entropies in terms of replica partition functions has enabled the introduction of various QMC estimators [\onlinecite{Melko2010:MutualInfo, Hastings2010:MeasureREE, Humeniuk2012:QMCREEgeneric, Inglis2013:WangLandau}].  While this has allowed for the investigation of many physical systems, the calculations remain costly due to the need to perform independent simulations either as a function of temperature or to incrementally compute the entanglement entropy for the region of interest.  These inefficiencies are compounded in two or more dimensions where the area law dominates and universal features are subtle.  Though improved estimators have been introduced [\onlinecite{Luitz2014:ImprovingEE},\onlinecite{Kulchytskyy2015:DetectingGoldstone}], these represent either incremental advancements or are restricted to specific models.  Simulations in this realm have yet to reach the large scale typically enjoyed by other more common QMC measurements.

At the same time, the ability to view the R\'{e}nyi entanglement entropy as a difference of free energies begs the question of whether one can apply nonequilibrium work relations [\onlinecite{Jarzynski1997:Noneql}, \onlinecite{Crooks1999:Noneql}] that have been widely used in the molecular dynamics community [\onlinecite{Ritort2006:MolecExp}].  Indeed, this idea was recently explored in the context of classical path integral Monte Carlo [\onlinecite{Alba2017:Jarzynski}], paving the way for the extension to QMC that we present here.  

We will first present the basic idea for a fixed subsystem size using replica partition functions.  We then show how this leads to the idea of smooth, space dependent quench functions that generalize the R\'{e}nyi entanglement entropy while preserving its universal features and allows for computations as a function of subsystem size.  These are used to extract universal features of SU($N$) magnets in one and two dimensions.  Finally we extend the method to $T=0$ projector QMC simulations in the valence bond basis [\onlinecite{Sandvik2005:GSproj}, \onlinecite{Sandvik2010:LoopVBB}], enabling ground state entanglement entropy measurements of the 2D Heisenberg model on truly large lattices.

{\em General method:}
The R\'{e}nyi entanglement entropy is given by
\begin{equation}
\label{eq:Srho}
S^{(n)}_A = \frac{1}{1-n}\ln(\mathrm{Tr}\rho^n_A),
\end{equation}
where $n$ is the R\'{e}nyi index ($n=2$ throughout this work) and $\rho_A=\mathrm{Tr}_{\bar{A}}\rho$ is the reduced density matrix of a subsystem $A$.  We express the density matrix as $\rho=e^{-\beta H}/Z$ with the partition function $Z=\mathrm{Tr}e^{-\beta H}$.  In what follows $\beta$ will be made sufficiently large so that only the ground state contributes. 

$S^{(n)}_A$ can be conveniently re-expressed as [\onlinecite{Calabrese2004:EEinQFT}]
\begin{equation}
\label{eq:renyi2}
S^{(n)}_A = \frac{1}{1-n}\ln\left(\frac{Z^{(n)}_{A}}{Z^{(n)}_{\o}}\right),
\end{equation}
where the replica partition functions are defined as
$Z^{(n)}_{A}= \mathrm{Tr} \left(\left( \mathrm{Tr}_{\bar{A}} e^{-\beta H}\right)^n\right) $ and 
$Z^{(n)}_{\o}= \left( \mathrm{Tr} e^{-\beta H}\right)^n =Z^n$.  Notice that $Z^{(n)}_{A}$
and $Z^{(n)}_{\o}$ are defined in exactly the same way, where $\o$ means that there are no sites in the $A$
subsystem.  We now wish to define a function $\mathcal{Z}^{(n)}_{A}(\lambda)$ such that $\mathcal{Z}^{(n)}_{A}(0)=Z^{(n)}_{\o}$ and  $\mathcal{Z}^{(n)}_{A}(1)=Z^{(n)}_{A}$.  This can be written explicitly as
\begin{equation}
\label{eq:Zlam}
\mathcal{Z}^{(n)}_{A}(\lambda) = \sum_{B \subseteq A} \lambda^{N_B}(1-\lambda)^{N_{A}-N_B}Z^{(n)}_B,
\end{equation}
where here $B$ denotes all possible subsets of $A$ including the empty set $\o$ and $A$ itself.  $N_{A}$
and $N_B$ denote the number of sites contained in the sets $A$ and $B$, respectively.  Here $\lambda$ controls the probability for an individual site in the $A$ subsystem to be traced only once.  To ease the notation in what follows we define $g_A(\lambda,N_B)=\lambda^{N_B}(1-\lambda)^{N_{A}-N_B}$.

The entanglement entropy can then be computed as:
\begin{equation}
\label{eq:Sdlam}
S^{(n)}_A = \frac{1}{1-n}\int^{1}_{0} d\lambda\frac{\partial \ln \mathcal{Z}^{(n)}_{A}(\lambda)}{\partial \lambda}.
\end{equation}
If one wanted to numerically compute this quantity using quantum Monte Carlo techniques, independent equilibrium simulations on a fine
grid of points between $\lambda=0$ and $\lambda=1$ would need to be performed.  Each simulation
would measure the equilibrium average $ \partial \ln \mathcal{Z}^{(n)}_{A}(\lambda)/\partial \lambda = \langle \frac{N_B}{\lambda} - \frac{N_A - N_B}{1-\lambda} \rangle_{\lambda}$
and the resulting curve from all the simulations would need to be numerically integrated.

This approach has obvious practical limitations.  It is much more desirable to estimate the entanglement
entropy directly from a single simulation.  Fortunately, this can be done by defining a nonequilibrium
process in which $\lambda$ is varied smoothly from 0 to 1:
\begin{equation}
\label{eq:Wd}
W^{(n)}_A = -\frac{1}{\beta}\int^{t_f}_{t_i} dt \frac{d\lambda}{dt}\frac{\partial \ln g_{A}(\lambda(t),N_B(t))}{\partial \lambda}
\end{equation}
Here $\lambda(t_i)=0$,  $\lambda(t_f)=1$ and $W^{(n)}_A$ is the total work done throughout the process.
We emphasize here that the work is a random variable that will follow a distribution.
One instance of the work is given by the sum of the increments $\partial \ln g_A(\lambda(t),N_B(t))$ along
a nonequilibrium path in the configuration space of $\mathcal{Z}^{(n)}_{A}(\lambda)$ as $\lambda$ is varied from 0 to 1.
The average  $\beta \langle W^{(n)}_A \rangle/(n-1)$ then approaches $S^{(n)}_A$ as $t_f - t_i \to \infty$.  If the quench time is
finite then the average will \emph{overestimate} $S^{(n)}_A$ due to the nonequilibrium entropy production associated with irreversibility.

Again the situation is undesirable, since we would like to accurately estimate the entanglement entropy even when the quench time is finite.  Here we can greatly prosper from a well known nonequilibrium fluctuation theorem of work, Jarzynski's equality [\onlinecite{Jarzynski1997:Noneql}], which in this
context reads:
\begin{equation}
\label{eq:jarzynski}
S^{(n)}_A=\frac{1}{1-n}\ln\left( \Big\langle e^{-\beta W^{(n)}_A} \Big\rangle\right),
\end{equation}
Remarkably this equality holds true regardless of the quench time.

\begin{figure}[!t]
\centerline{\includegraphics[angle=0,width=1.0\columnwidth]{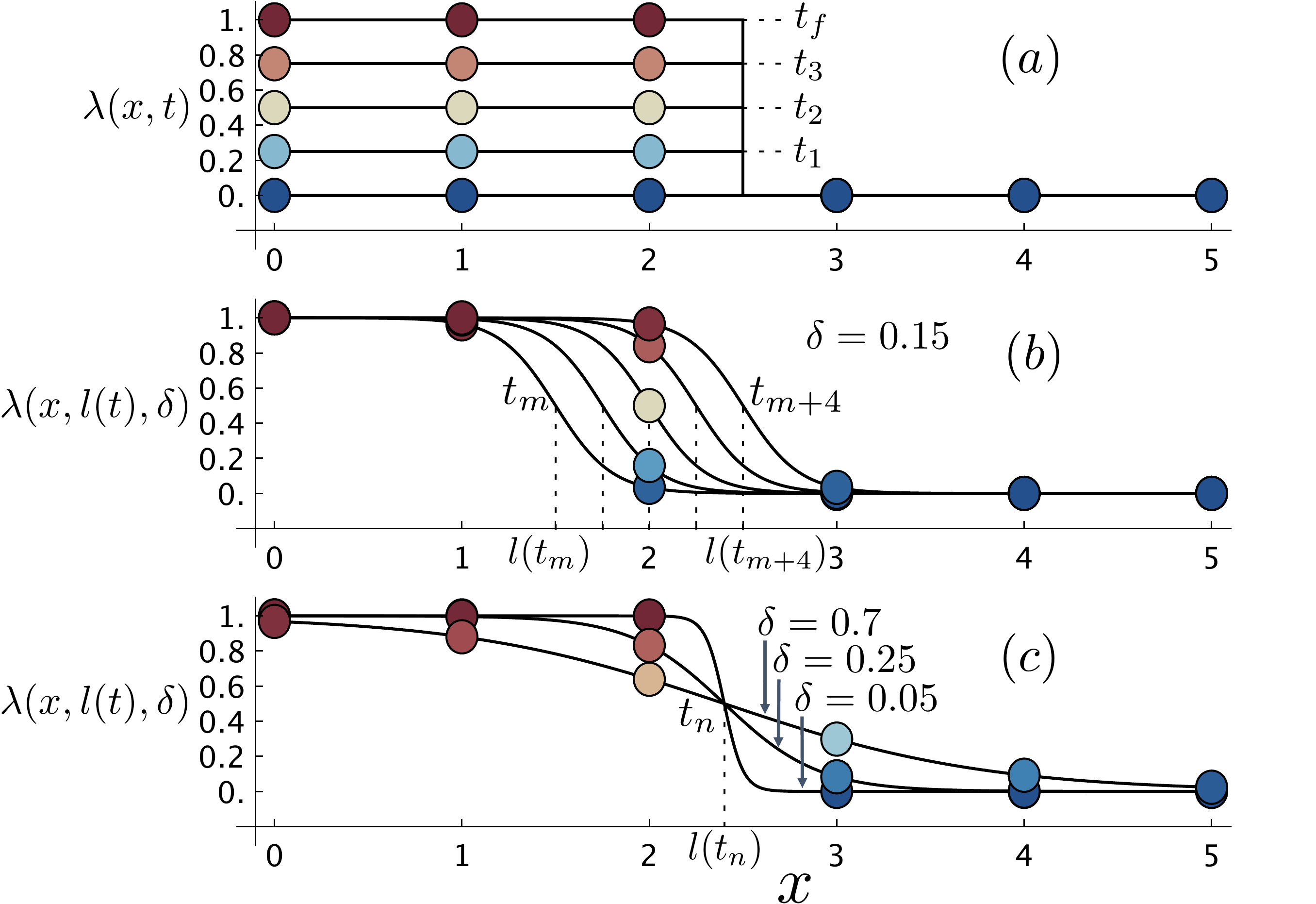}}
\caption{Panel (a): The quench protocol outlined in the general method section.  The external field $\lambda$ is taken to be spatially constant in the $A$ subsystem (here three sites of a six site chain) and varying in time from 0 to 1.  Panel (b): The quench function in Eq. (\ref{eq:lambda_l}) shown at five different time steps.  Here the site at $x=2$ is smoothly brought into the entangling region as the subsystem partition is moved from left to right.  Panel (c): The same quench function at one instant in time shown for several values of $\delta$.  When $\delta$ is small each site is quenched individually, computing exactly the R\'enyi entanglement entropy of a block subsystem.  For larger $\delta$ the subsystem boundary is spread over several sites, generalizing the R\'enyi entanglement entropy in such a way that lattice scale features are suppressed while universal information is preserved.}
\label{fig:qpchain1}
\end{figure}

\begin{figure}[h],
\centerline{\includegraphics[angle=0,width=1.0\columnwidth]{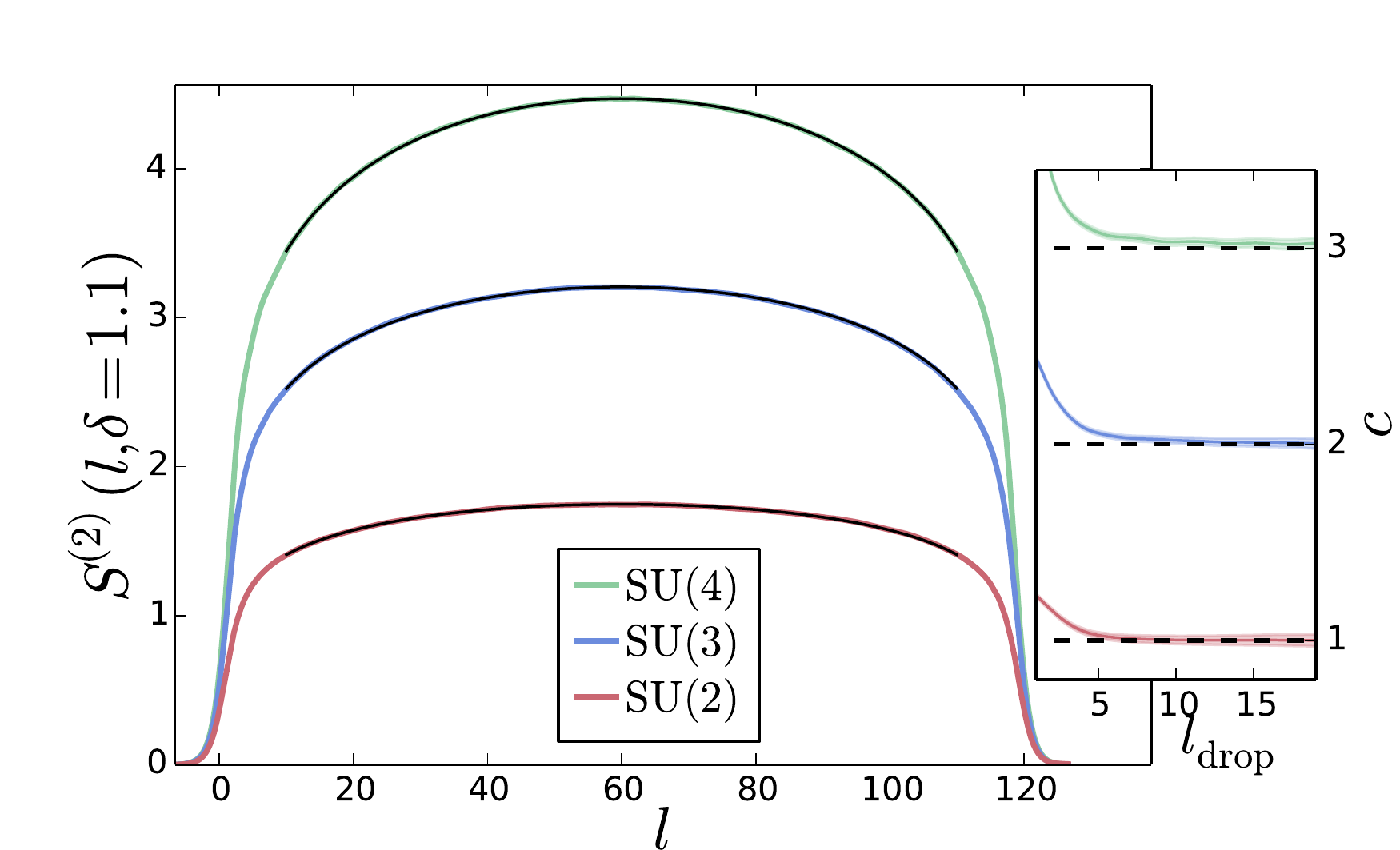}}
\caption{Here we compute the R\'enyi entanglement entropy using the quench function in Eq. (\ref{eq:lambda_l}) for an $L=120$ chain for $\mathrm{SU}(2)$, $\mathrm{SU}(3)$ and $\mathrm{SU}(4)$ with $\delta=1.1$.  The colored curves with shaded error bar are the QMC data and black lines are numerical fits.  We find perfect agreement with the universal form in Eq. (\ref{eq:LogScaling}) with central charge $c=N-1$.}
\label{fig:L120su234}
\end{figure}

{\em Quench protocols:}
In the previous section we took the external field $\lambda$ to be constant in space and varying in time between 0 and 1. This is depicted in panel $(a)$ of Fig. \ref{fig:qpchain1}.  This provides us with a very efficient way of computing the entanglement entropy of large regions, which we will demonstrate in the final section using $T=0$ projector QMC simulations.  At this point, however, we would like to introduce another quench protocol that can compute the entanglement entropy as a function of subsystem size in a single nonequilibrium simulation.

This can be achieved by making $\lambda$ a function of both time and space in such a way that the subsystem partition ``slides" across the lattice, in which case Eq. (\ref{eq:Zlam}) can be easily adapted (see supplemental material).  This naturally leads us to the idea that $\lambda$ should be a smooth function of space so that individual sites are quenched gradually while the partition moves.
A convenient choice is given by
\begin{equation}
\label{eq:lambda_l}
\lambda(x,l(t),\delta)=\frac{1}{1+e^{(x-l(t)+\frac{1}{2})/\delta}}.
\end{equation}
Here $x$ is the spatial coordinate (from 0 to $L-1$) along the chain and $\lambda$ depends on time through the parameter $l(t)$ which represents the center of one boundary between the $A$ subsystem and the rest of the chain (see panel $(b)$ and $(c)$ of Fig. \ref{fig:qpchain1}).  A shift of $\tfrac{1}{2}$ is introduced so that $l(t)$ is centered between sites of the chain.  We have also introduced a parameter $\delta$ that controls the width of the boundary.  As we will discuss momentarily, $\delta$ allows us to suppress lattice scale features of the entanglement entropy while preserving the universal features.

The quench protocol as defined by Eq (\ref{eq:lambda_l}) involves varying $l(t)$ linearly in time from $l(t_i)=-p\delta$ to $l(t_f)=L-1+p\delta$, where the constant $p$ is used to ensure that $\lambda(x,l(t_i),\delta)$ and $\lambda(x,l(t_f),\delta)$ are approximately 0 and 1 everywhere, respectively.  As $l(t)$ is varied along the length of the chain, the total work is computed as a function of time.  Since $l(t)$ represents the current location of the subsystem boundary, this allows us to compute the entanglement entropy  as a function of subsystem size in a single nonequilibrium simulation.  Furthermore, since $l(t)$ is real valued, the entanglement entropy is computed as a continuous curve that can be sampled up to the number of time steps used in the nonequilibrium process.

We now wish to discuss the role of the parameter $\delta$, which is a central idea that emerges from our new method. When $\delta$ is small enough so that the width of the subsystem boundary is much less than the lattice spacing, our protocol allows for the exact computation of $S^{(2)}_A$ as a function of subsystem size.  The entanglement entropy thus defined contains strong lattice-scale features that can often obscure universal properties that one wishes to extract.  By increasing $\delta$ so that the subsystem partition is spread over a few lattice sites we effectively average over features below a certain scale.  As we intuitively expect, the presence of $\delta$ will only affect non-universal features of the entanglement entropy, namely the area law contribution.

{\em SU(N) chain:}
As a first test case for our new method, we start with the SU($N$) Heisenberg antiferromagnet on a periodic one dimensional chain [\onlinecite{Sutherland1975:Multi}, \onlinecite{Affleck1988:CriticalSUn}].  The Hamiltonian can be written simply as a nearest neighbor permutation:
\begin{equation}
\label{eq:suNchainham}
H=\frac{J}{N}\sum_{i}\sum^{N}_{\alpha, \beta =1} |\alpha_i \beta_{i+1} \rangle \langle \beta_{i} \alpha_{i+1}| .
\end{equation}
The ground state of this model is critical, and belongs to the family of conformally invariant Wess-Zumino-Witten nonlinear sigma models with central charge $c=N-1$  [\onlinecite{Affleck1988:CriticalSUn}].  This system gives rise to the celebrated log violation of the area law, which is given by the following form:
\begin{equation}
\label{eq:LogScaling}
S^{(n)}(l)=\frac{c}{6}\left(1+\frac{1}{n}\right) \log \left( \frac{L}{\pi} \sin \left( \frac{\pi l}{L} \right) \right) + b.
\end{equation}
The R\'{e}nyi entanglement entropies of this system also show oscillations as a function of subsystem size [\onlinecite{Demidio2015:Renyi}].  In Fig. \ref{fig:L120su234} we use our newly developed nonequilibrium method in the framework of the stochastic series expansion QMC algorithm [\onlinecite{Sandvik2010:CompStud}], combined the with quench protocol in Eqn. (\ref{eq:lambda_l}) to show that lattice scale oscillations of the entanglement entropy can be suppressed by smoothing out the subsystem boundary using the parameter $\delta$.  Crucially the central charge is insensitive to the presence of $\delta$, which only affects the nonuniversal constant (area law) contribution (see supplemental materials).  As expected, we find perfect agreement with the central charge $c=N-1$.

\begin{figure}[!t]
\centerline{\includegraphics[angle=0,width=1.0\columnwidth]{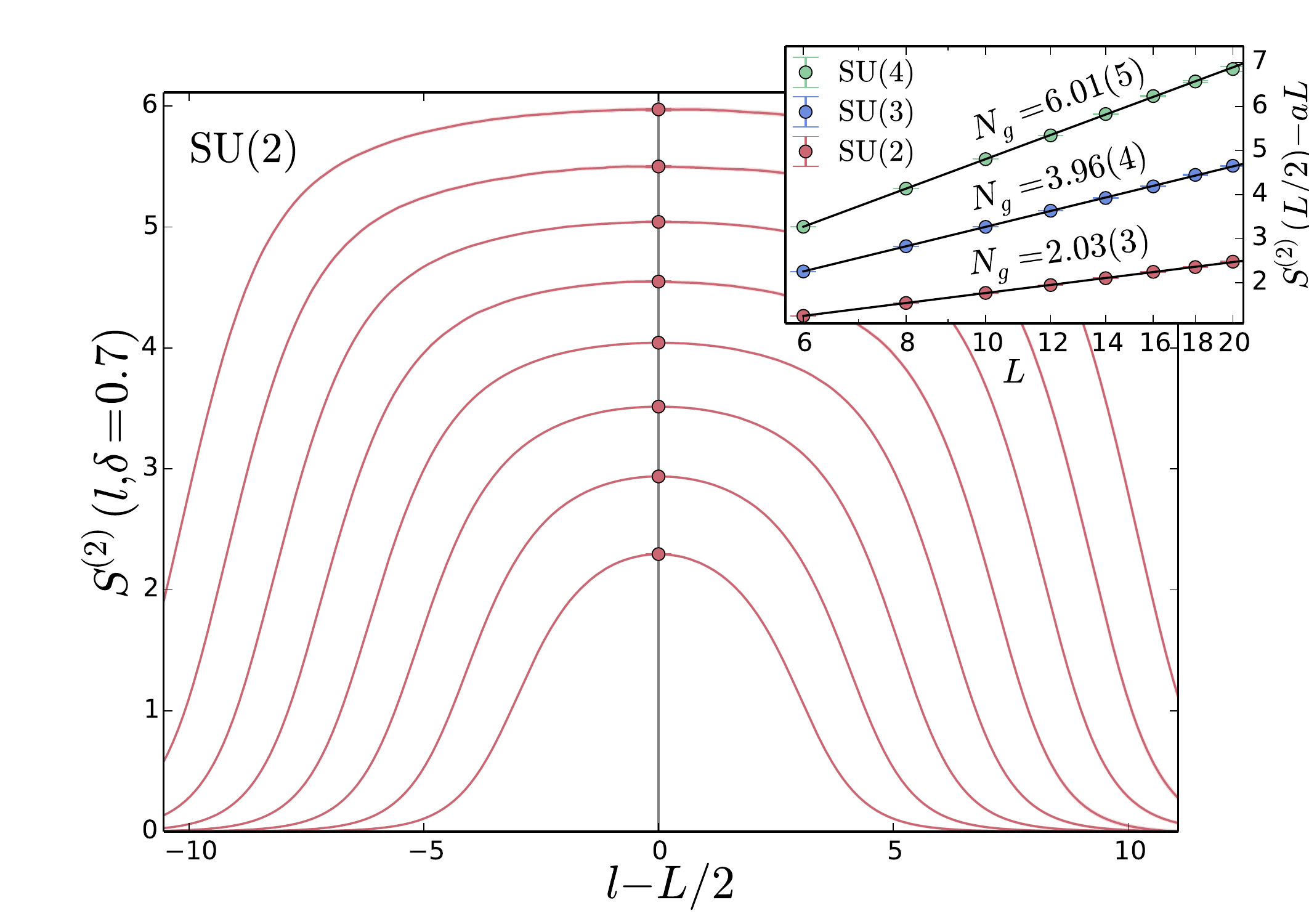}}
\caption{In the main plot we show the data collected using our two dimensional version of the quench function in Eq. (\ref{eq:lambda_l}) on the staggered $\mathrm{SU}(2)$ model for  $L=6,8,10,..., 20$, $J_2/J_1=1$ and $\delta=0.7$.  We use the data from the center cut to fit to the universal scaling form in Eq. (\ref{eq:goldlog}) and in the inset we plot the center cut data with the area law piece subtracted on semi-log axes along with the best fit (black line).  This procedure is repeated for $\mathrm{SU}(3)$ and $\mathrm{SU}(4)$ (also shown in the inset) with $J_2/J_1=2$ and $J_2/J_1=3.5$, respectively.  We find excellent agreement with the number of Goldstone modes $N_g=2(N-1)$.}
\label{fig:2dsu2L16}
\end{figure}

{\em 2D staggered SU(N) Model:} 
Recently, the entanglement entropy has been used to detect Goldstone modes in magnetically ordered systems [\onlinecite{Metlitski2011:GoldstoneEE}, \onlinecite{Kulchytskyy2015:DetectingGoldstone}].  To this end, we select a family of two dimensional SU($N$) symmetric models that support gapless spin wave excitations.  The Hamiltonian is given by [\onlinecite{Kaul2013:Bridging}]
\begin{equation}
\label{eq:2DHam}
H=-\frac{J_1}{N}\sum_{\langle ij \rangle}\sum^{N}_{\alpha, \beta =1} |\alpha_i \alpha_j \rangle \langle \beta_i \beta_j| -\frac{J_2}{N}\sum_{\langle \langle ij \rangle \rangle}\sum^{N}_{\alpha, \beta =1} |\alpha_i \beta_j \rangle \langle \beta_i \alpha_j| .
\end{equation}
Here we consider periodic square lattices, using the fundamental representation of SU($N$) on one sublattice and the conjugate to the fundamental on the other sublattice.  $J_1>0$ and $J_2>0$ are the nearest and next-nearest neighbor couplings, respectively. 

This model is an SU($N$) generalization of the spin-$\tfrac{1}{2}$ Heisenberg antiferromagnet with a \textit{ferromagnetic} next-nearest neighbor interaction.  When $J_2=0$ the ground state is magnetically ordered with gapless Goldstone modes for $2\leq N \leq 4$ and forms an valence bond solid for $N \geq 5$ [\onlinecite{Kaul2012:LargeN}].  When $J_2 > 0$ magnetic order is enhanced which greatly facilitates the ability to accurately extract the number of Goldstone modes by fitting to our entanglement entropy data.

The presence of Goldstone modes produces a log contribution to the entanglement entropy [\onlinecite{Metlitski2011:GoldstoneEE}].  In the simplest case, we can take an $L$ x $L$ cylinder and cut it in half.  The resulting entanglement entropy as a function of the linear system size $L$ is given by
\begin{equation}
\label{eq:goldlog}
S^{(n)}(L/2)=a L+\frac{N_g}{2} \log (L) + b.
\end{equation}
Here $N_g$ is the number of Goldstone modes and b contains a geometrical constant.  

We again apply our nonequilibrium method in combination with the quench protocol in Eq. (\ref{eq:lambda_l}), where now the quench function is taken to be spatially constant in the $y$-direction and the partition is swept along the $x$-direction.  Fig. (\ref{fig:2dsu2L16}) shows the resulting curves for system sizes $L=6,8,..,20$ for SU(2).  We take the midpoint of our data and perform a fit to Eq. (\ref{eq:goldlog}).  This procedure is performed for SU(2), SU(3) and SU(4) with $J_2/J_1=1,2,3.5$, respectively.  The inset of Fig. (\ref{fig:2dsu2L16}) shows the entanglement entropy at the center cut with the area law piece subtracted.  On semi-log axes, the log contribution manifests itself as a straight line.  Within one percent accuracy we find the number of Goldstone modes to be $N_g=2,4,6$ for $N=2,3,4$ respectively.

\begin{figure}[!h]
\centerline{\includegraphics[angle=0,width=1.0\columnwidth]{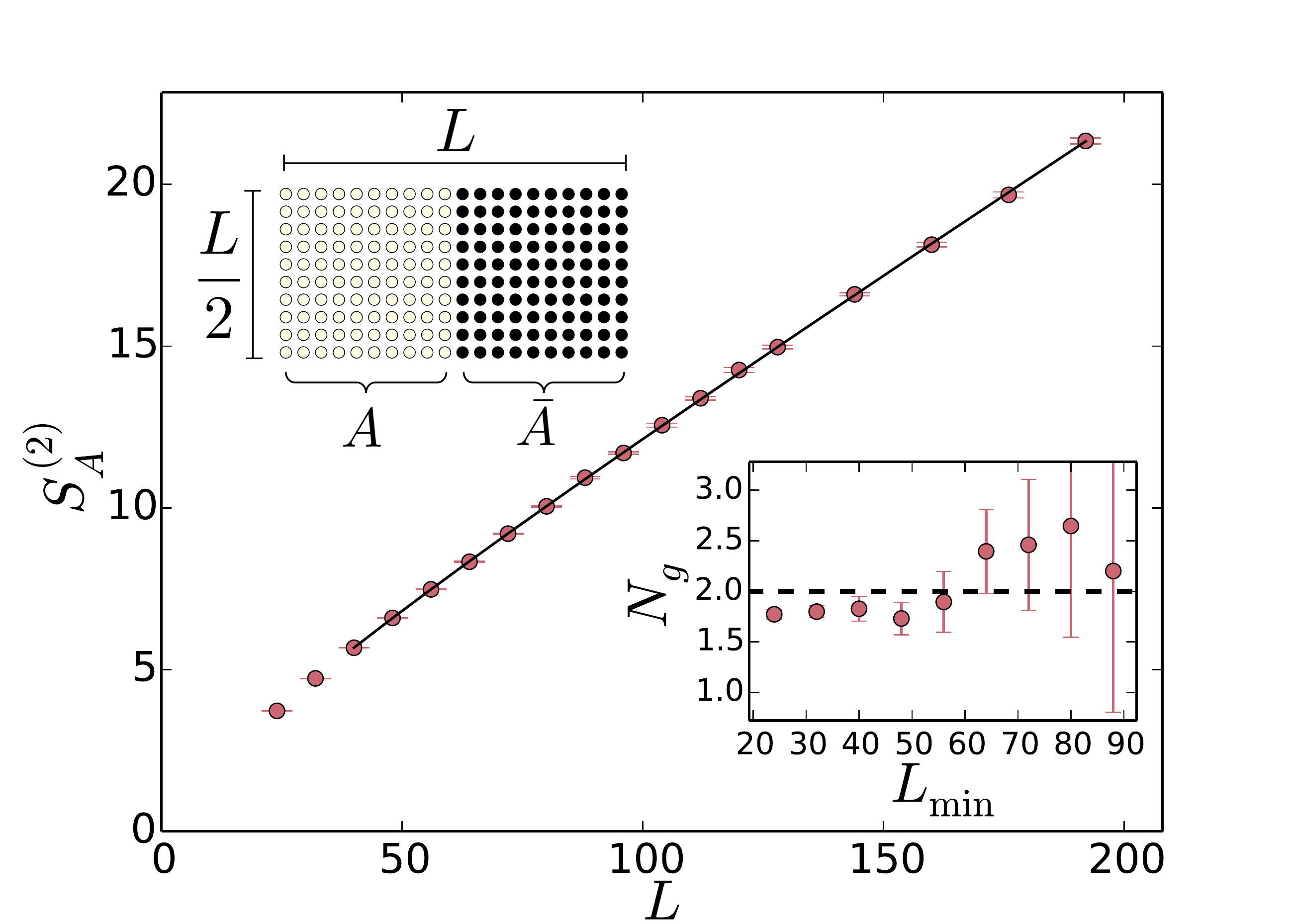}}
\caption{Half-system second R\'{e}nyi entanglement entropy for the spin-$\tfrac{1}{2}$ square lattice Heisenberg model.  By using $T=0$ projector QMC simulations in the valence bond basis we can much more efficiently reach the ground state.  This allows for precise calculations on unprecedented system sizes (up to size 192 x 96 shown here).  Without adding a next-nearest-neighbor coupling, we observe very slow convergence to $N_g=2$ when fitting to the scaling form (\ref{eq:goldlog}).}
\label{fig:T0proj}
\end{figure}

{\em Projector QMC simulations:} 
We finally wish to extend our method to $T=0$ projector QMC simulations in the valence bond basis [\onlinecite{Sandvik2005:GSproj}, \onlinecite{Sandvik2010:LoopVBB}].  The motivation here is that one can much more easily reach the groundstate as compared with the finite temperature stochastic series expansion algorithm.  Furthermore, we directly compute the half-system second R\'{e}nyi entanglement entropy by performing the quench analogous to panel (a) of Fig. \ref{fig:qpchain1}.  These efficiency gains allow us to access unprecedented system sizes for the square lattice spin-$\tfrac{1}{2}$ Heisenberg model, shown in Fig. \ref{fig:T0proj}.  Finally, we note that the convergence of $N_g$ by fitting to the form in Eq. (\ref{eq:goldlog}) is very slow when a next-nearest-neighbor coupling is absent.

{\em Conclusion:} 
We have introduced a highly efficient nonequilibrium method for computing the R\'{e}nyi entanglement entropy in the context of statistically exact QMC simulations.  Remarkably, in all of our simulations we observe that the QMC statistical error remains constant when the number of measurement sweeps is proportional to the number of sites in the entangling region (see supplemental material).  Room for improvement is also still possible since a linear quench protocol for a fixed region is most certainly not optimal.

{\em Acknowledgements:}  We thank Fr\'ed\'eric Mila for his encouragement and support for this project.  We also thank Ribhu K. Kaul for suggesting the 2D staggered SU(N) model as a test case for this method.  SSE simulations were carried out on the Fidis cluster at EPFL and projector simulations on Comet at SDSC with the Extreme Science and Engineering Discovery Environment (XSEDE) Allocation No. TG- DMR130040.

{\em Note added:}
Recently, another important work on com- puting entanglement entropy using thermodynamic concepts via a field theoretic mapping has also been introduced [\onlinecite{Santos2020:VNnoWave}].

\bibliography{NonEqlEE}

\begin{thebibliography}{37}%
\makeatletter
\providecommand \@ifxundefined [1]{%
 \@ifx{#1\undefined}
}%
\providecommand \@ifnum [1]{%
 \ifnum #1\expandafter \@firstoftwo
 \else \expandafter \@secondoftwo
 \fi
}%
\providecommand \@ifx [1]{%
 \ifx #1\expandafter \@firstoftwo
 \else \expandafter \@secondoftwo
 \fi
}%
\providecommand \natexlab [1]{#1}%
\providecommand \enquote  [1]{``#1''}%
\providecommand \bibnamefont  [1]{#1}%
\providecommand \bibfnamefont [1]{#1}%
\providecommand \citenamefont [1]{#1}%
\providecommand \href@noop [0]{\@secondoftwo}%
\providecommand \href [0]{\begingroup \@sanitize@url \@href}%
\providecommand \@href[1]{\@@startlink{#1}\@@href}%
\providecommand \@@href[1]{\endgroup#1\@@endlink}%
\providecommand \@sanitize@url [0]{\catcode `\\12\catcode `\$12\catcode
  `\&12\catcode `\#12\catcode `\^12\catcode `\_12\catcode `\%12\relax}%
\providecommand \@@startlink[1]{}%
\providecommand \@@endlink[0]{}%
\providecommand \url  [0]{\begingroup\@sanitize@url \@url }%
\providecommand \@url [1]{\endgroup\@href {#1}{\urlprefix }}%
\providecommand \urlprefix  [0]{URL }%
\providecommand \Eprint [0]{\href }%
\providecommand \doibase [0]{http://dx.doi.org/}%
\providecommand \selectlanguage [0]{\@gobble}%
\providecommand \bibinfo  [0]{\@secondoftwo}%
\providecommand \bibfield  [0]{\@secondoftwo}%
\providecommand \translation [1]{[#1]}%
\providecommand \BibitemOpen [0]{}%
\providecommand \bibitemStop [0]{}%
\providecommand \bibitemNoStop [0]{.\EOS\space}%
\providecommand \EOS [0]{\spacefactor3000\relax}%
\providecommand \BibitemShut  [1]{\csname bibitem#1\endcsname}%
\let\auto@bib@innerbib\@empty
\bibitem [{\citenamefont {Amico}\ \emph {et~al.}(2008)\citenamefont {Amico},
  \citenamefont {Fazio}, \citenamefont {Osterloh},\ and\ \citenamefont
  {Vedral}}]{Amico2008:EE}%
  \BibitemOpen
  \bibfield  {author} {\bibinfo {author} {\bibfnamefont {Luigi}\ \bibnamefont
  {Amico}}, \bibinfo {author} {\bibfnamefont {Rosario}\ \bibnamefont {Fazio}},
  \bibinfo {author} {\bibfnamefont {Andreas}\ \bibnamefont {Osterloh}}, \ and\
  \bibinfo {author} {\bibfnamefont {Vlatko}\ \bibnamefont {Vedral}},\
  }\bibfield  {title} {\enquote {\bibinfo {title} {Entanglement in many-body
  systems},}\ }\href {\doibase 10.1103/RevModPhys.80.517} {\bibfield  {journal}
  {\bibinfo  {journal} {Rev. Mod. Phys.}\ }\textbf {\bibinfo {volume} {80}},\
  \bibinfo {pages} {517--576} (\bibinfo {year} {2008})}\BibitemShut {NoStop}%
\bibitem [{\citenamefont {Eisert}\ \emph {et~al.}(2010)\citenamefont {Eisert},
  \citenamefont {Cramer},\ and\ \citenamefont {Plenio}}]{Eisert2010:AreaLaws}%
  \BibitemOpen
  \bibfield  {author} {\bibinfo {author} {\bibfnamefont {J.}~\bibnamefont
  {Eisert}}, \bibinfo {author} {\bibfnamefont {M.}~\bibnamefont {Cramer}}, \
  and\ \bibinfo {author} {\bibfnamefont {M.~B.}\ \bibnamefont {Plenio}},\
  }\bibfield  {title} {\enquote {\bibinfo {title} {Colloquium: Area laws for
  the entanglement entropy},}\ }\href {\doibase 10.1103/RevModPhys.82.277}
  {\bibfield  {journal} {\bibinfo  {journal} {Rev. Mod. Phys.}\ }\textbf
  {\bibinfo {volume} {82}},\ \bibinfo {pages} {277--306} (\bibinfo {year}
  {2010})}\BibitemShut {NoStop}%
\bibitem [{\citenamefont {Laflorencie}(2016)}]{Laflorencie2016:Entanglement}%
  \BibitemOpen
  \bibfield  {author} {\bibinfo {author} {\bibfnamefont {Nicolas}\ \bibnamefont
  {Laflorencie}},\ }\bibfield  {title} {\enquote {\bibinfo {title} {Quantum
  entanglement in condensed matter systems},}\ }\href {\doibase
  https://doi.org/10.1016/j.physrep.2016.06.008} {\bibfield  {journal}
  {\bibinfo  {journal} {Physics Reports}\ }\textbf {\bibinfo {volume} {646}},\
  \bibinfo {pages} {1--59} (\bibinfo {year} {2016})}\BibitemShut {NoStop}%
\bibitem [{\citenamefont {Holzhey}\ \emph {et~al.}(1994)\citenamefont
  {Holzhey}, \citenamefont {Larsen},\ and\ \citenamefont
  {Wilczek}}]{Holzhey1994:Geometric}%
  \BibitemOpen
  \bibfield  {author} {\bibinfo {author} {\bibfnamefont {Christoph}\
  \bibnamefont {Holzhey}}, \bibinfo {author} {\bibfnamefont {Finn}\
  \bibnamefont {Larsen}}, \ and\ \bibinfo {author} {\bibfnamefont {Frank}\
  \bibnamefont {Wilczek}},\ }\bibfield  {title} {\enquote {\bibinfo {title}
  {Geometric and renormalized entropy in conformal field theory},}\ }\href
  {\doibase https://doi.org/10.1016/0550-3213(94)90402-2} {\bibfield  {journal}
  {\bibinfo  {journal} {Nuclear Physics B}\ }\textbf {\bibinfo {volume}
  {424}},\ \bibinfo {pages} {443 -- 467} (\bibinfo {year} {1994})}\BibitemShut
  {NoStop}%
\bibitem [{\citenamefont {Vidal}\ \emph {et~al.}(2003)\citenamefont {Vidal},
  \citenamefont {Latorre}, \citenamefont {Rico},\ and\ \citenamefont
  {Kitaev}}]{Vidal2003:EinQCP}%
  \BibitemOpen
  \bibfield  {author} {\bibinfo {author} {\bibfnamefont {G.}~\bibnamefont
  {Vidal}}, \bibinfo {author} {\bibfnamefont {J.~I.}\ \bibnamefont {Latorre}},
  \bibinfo {author} {\bibfnamefont {E.}~\bibnamefont {Rico}}, \ and\ \bibinfo
  {author} {\bibfnamefont {A.}~\bibnamefont {Kitaev}},\ }\bibfield  {title}
  {\enquote {\bibinfo {title} {Entanglement in quantum critical phenomena},}\
  }\href {\doibase 10.1103/PhysRevLett.90.227902} {\bibfield  {journal}
  {\bibinfo  {journal} {Phys. Rev. Lett.}\ }\textbf {\bibinfo {volume} {90}},\
  \bibinfo {pages} {227902} (\bibinfo {year} {2003})}\BibitemShut {NoStop}%
\bibitem [{\citenamefont {Calabrese}\ and\ \citenamefont
  {Cardy}(2004)}]{Calabrese2004:EEinQFT}%
  \BibitemOpen
  \bibfield  {author} {\bibinfo {author} {\bibfnamefont {Pasquale}\
  \bibnamefont {Calabrese}}\ and\ \bibinfo {author} {\bibfnamefont {John}\
  \bibnamefont {Cardy}},\ }\bibfield  {title} {\enquote {\bibinfo {title}
  {Entanglement entropy and quantum field theory},}\ }\href {\doibase
  10.1088/1742-5468/2004/06/p06002} {\bibfield  {journal} {\bibinfo  {journal}
  {Journal of Statistical Mechanics: Theory and Experiment}\ }\textbf {\bibinfo
  {volume} {2004}},\ \bibinfo {pages} {P06002} (\bibinfo {year}
  {2004})}\BibitemShut {NoStop}%
\bibitem [{\citenamefont {{Metlitski}}\ and\ \citenamefont
  {{Grover}}(2011)}]{Metlitski2011:GoldstoneEE}%
  \BibitemOpen
  \bibfield  {author} {\bibinfo {author} {\bibfnamefont {Max~A.}\ \bibnamefont
  {{Metlitski}}}\ and\ \bibinfo {author} {\bibfnamefont {Tarun}\ \bibnamefont
  {{Grover}}},\ }\bibfield  {title} {\enquote {\bibinfo {title} {{Entanglement
  Entropy of Systems with Spontaneously Broken Continuous Symmetry}},}\
  }\href@noop {} {\bibfield  {journal} {\bibinfo  {journal} {arXiv e-prints}\
  ,\ \bibinfo {eid} {arXiv:1112.5166}} (\bibinfo {year} {2011})},\ \Eprint
  {http://arxiv.org/abs/1112.5166} {arXiv:1112.5166 [cond-mat.str-el]}
  \BibitemShut {NoStop}%
\bibitem [{\citenamefont {Castro-Alvaredo}\ and\ \citenamefont
  {Doyon}(2012)}]{Castro-Alvaredo2012:EEhighDeg}%
  \BibitemOpen
  \bibfield  {author} {\bibinfo {author} {\bibfnamefont {Olalla~A.}\
  \bibnamefont {Castro-Alvaredo}}\ and\ \bibinfo {author} {\bibfnamefont
  {Benjamin}\ \bibnamefont {Doyon}},\ }\bibfield  {title} {\enquote {\bibinfo
  {title} {Entanglement entropy of highly degenerate states and fractal
  dimensions},}\ }\href {\doibase 10.1103/PhysRevLett.108.120401} {\bibfield
  {journal} {\bibinfo  {journal} {Phys. Rev. Lett.}\ }\textbf {\bibinfo
  {volume} {108}},\ \bibinfo {pages} {120401} (\bibinfo {year}
  {2012})}\BibitemShut {NoStop}%
\bibitem [{\citenamefont {Song}\ \emph {et~al.}(2011)\citenamefont {Song},
  \citenamefont {Laflorencie}, \citenamefont {Rachel},\ and\ \citenamefont
  {Le~Hur}}]{Song2011:EEHeisen}%
  \BibitemOpen
  \bibfield  {author} {\bibinfo {author} {\bibfnamefont {H.~Francis}\
  \bibnamefont {Song}}, \bibinfo {author} {\bibfnamefont {Nicolas}\
  \bibnamefont {Laflorencie}}, \bibinfo {author} {\bibfnamefont {Stephan}\
  \bibnamefont {Rachel}}, \ and\ \bibinfo {author} {\bibfnamefont {Karyn}\
  \bibnamefont {Le~Hur}},\ }\bibfield  {title} {\enquote {\bibinfo {title}
  {Entanglement entropy of the two-dimensional heisenberg antiferromagnet},}\
  }\href {\doibase 10.1103/PhysRevB.83.224410} {\bibfield  {journal} {\bibinfo
  {journal} {Phys. Rev. B}\ }\textbf {\bibinfo {volume} {83}},\ \bibinfo
  {pages} {224410} (\bibinfo {year} {2011})}\BibitemShut {NoStop}%
\bibitem [{\citenamefont {Ju}\ \emph {et~al.}(2012)\citenamefont {Ju},
  \citenamefont {Kallin}, \citenamefont {Fendley}, \citenamefont {Hastings},\
  and\ \citenamefont {Melko}}]{Ju2012:2Dgapless}%
  \BibitemOpen
  \bibfield  {author} {\bibinfo {author} {\bibfnamefont {Hyejin}\ \bibnamefont
  {Ju}}, \bibinfo {author} {\bibfnamefont {Ann~B.}\ \bibnamefont {Kallin}},
  \bibinfo {author} {\bibfnamefont {Paul}\ \bibnamefont {Fendley}}, \bibinfo
  {author} {\bibfnamefont {Matthew~B.}\ \bibnamefont {Hastings}}, \ and\
  \bibinfo {author} {\bibfnamefont {Roger~G.}\ \bibnamefont {Melko}},\
  }\bibfield  {title} {\enquote {\bibinfo {title} {Entanglement scaling in
  two-dimensional gapless systems},}\ }\href {\doibase
  10.1103/PhysRevB.85.165121} {\bibfield  {journal} {\bibinfo  {journal} {Phys.
  Rev. B}\ }\textbf {\bibinfo {volume} {85}},\ \bibinfo {pages} {165121}
  (\bibinfo {year} {2012})}\BibitemShut {NoStop}%
\bibitem [{\citenamefont {Luitz}\ \emph {et~al.}(2015)\citenamefont {Luitz},
  \citenamefont {Plat}, \citenamefont {Alet},\ and\ \citenamefont
  {Laflorencie}}]{Luitz2015:UniversalLog}%
  \BibitemOpen
  \bibfield  {author} {\bibinfo {author} {\bibfnamefont {David~J.}\
  \bibnamefont {Luitz}}, \bibinfo {author} {\bibfnamefont {Xavier}\
  \bibnamefont {Plat}}, \bibinfo {author} {\bibfnamefont {Fabien}\ \bibnamefont
  {Alet}}, \ and\ \bibinfo {author} {\bibfnamefont {Nicolas}\ \bibnamefont
  {Laflorencie}},\ }\bibfield  {title} {\enquote {\bibinfo {title} {Universal
  logarithmic corrections to entanglement entropies in two dimensions with
  spontaneously broken continuous symmetries},}\ }\href {\doibase
  10.1103/PhysRevB.91.155145} {\bibfield  {journal} {\bibinfo  {journal} {Phys.
  Rev. B}\ }\textbf {\bibinfo {volume} {91}},\ \bibinfo {pages} {155145}
  (\bibinfo {year} {2015})}\BibitemShut {NoStop}%
\bibitem [{\citenamefont {Laflorencie}\ \emph {et~al.}(2015)\citenamefont
  {Laflorencie}, \citenamefont {Luitz},\ and\ \citenamefont
  {Alet}}]{Laflorencie2015:SpinWaveEE}%
  \BibitemOpen
  \bibfield  {author} {\bibinfo {author} {\bibfnamefont {Nicolas}\ \bibnamefont
  {Laflorencie}}, \bibinfo {author} {\bibfnamefont {David~J.}\ \bibnamefont
  {Luitz}}, \ and\ \bibinfo {author} {\bibfnamefont {Fabien}\ \bibnamefont
  {Alet}},\ }\bibfield  {title} {\enquote {\bibinfo {title} {Spin-wave approach
  for entanglement entropies of the ${J}_{1}\ensuremath{-}{J}_{2}$ heisenberg
  antiferromagnet on the square lattice},}\ }\href {\doibase
  10.1103/PhysRevB.92.115126} {\bibfield  {journal} {\bibinfo  {journal} {Phys.
  Rev. B}\ }\textbf {\bibinfo {volume} {92}},\ \bibinfo {pages} {115126}
  (\bibinfo {year} {2015})}\BibitemShut {NoStop}%
\bibitem [{\citenamefont {Rademaker}(2015)}]{Rademaker2015:TowerOfStates}%
  \BibitemOpen
  \bibfield  {author} {\bibinfo {author} {\bibfnamefont {Louk}\ \bibnamefont
  {Rademaker}},\ }\bibfield  {title} {\enquote {\bibinfo {title} {Tower of
  states and the entanglement spectrum in a coplanar antiferromagnet},}\ }\href
  {\doibase 10.1103/PhysRevB.92.144419} {\bibfield  {journal} {\bibinfo
  {journal} {Phys. Rev. B}\ }\textbf {\bibinfo {volume} {92}},\ \bibinfo
  {pages} {144419} (\bibinfo {year} {2015})}\BibitemShut {NoStop}%
\bibitem [{\citenamefont {Kallin}\ \emph {et~al.}(2011)\citenamefont {Kallin},
  \citenamefont {Hastings}, \citenamefont {Melko},\ and\ \citenamefont
  {Singh}}]{Kallin2011:AnomaliesEEHeisen}%
  \BibitemOpen
  \bibfield  {author} {\bibinfo {author} {\bibfnamefont {Ann~B.}\ \bibnamefont
  {Kallin}}, \bibinfo {author} {\bibfnamefont {Matthew~B.}\ \bibnamefont
  {Hastings}}, \bibinfo {author} {\bibfnamefont {Roger~G.}\ \bibnamefont
  {Melko}}, \ and\ \bibinfo {author} {\bibfnamefont {Rajiv R.~P.}\ \bibnamefont
  {Singh}},\ }\bibfield  {title} {\enquote {\bibinfo {title} {Anomalies in the
  entanglement properties of the square-lattice heisenberg model},}\ }\href
  {\doibase 10.1103/PhysRevB.84.165134} {\bibfield  {journal} {\bibinfo
  {journal} {Phys. Rev. B}\ }\textbf {\bibinfo {volume} {84}},\ \bibinfo
  {pages} {165134} (\bibinfo {year} {2011})}\BibitemShut {NoStop}%
\bibitem [{\citenamefont {Kulchytskyy}\ \emph {et~al.}(2015)\citenamefont
  {Kulchytskyy}, \citenamefont {Herdman}, \citenamefont {Inglis},\ and\
  \citenamefont {Melko}}]{Kulchytskyy2015:DetectingGoldstone}%
  \BibitemOpen
  \bibfield  {author} {\bibinfo {author} {\bibfnamefont {Bohdan}\ \bibnamefont
  {Kulchytskyy}}, \bibinfo {author} {\bibfnamefont {C.~M.}\ \bibnamefont
  {Herdman}}, \bibinfo {author} {\bibfnamefont {Stephen}\ \bibnamefont
  {Inglis}}, \ and\ \bibinfo {author} {\bibfnamefont {Roger~G.}\ \bibnamefont
  {Melko}},\ }\bibfield  {title} {\enquote {\bibinfo {title} {Detecting
  goldstone modes with entanglement entropy},}\ }\href {\doibase
  10.1103/PhysRevB.92.115146} {\bibfield  {journal} {\bibinfo  {journal} {Phys.
  Rev. B}\ }\textbf {\bibinfo {volume} {92}},\ \bibinfo {pages} {115146}
  (\bibinfo {year} {2015})}\BibitemShut {NoStop}%
\bibitem [{\citenamefont {Luitz}\ and\ \citenamefont
  {Laflorencie}(2017)}]{Luitz2017:QMClinesubsystem}%
  \BibitemOpen
  \bibfield  {author} {\bibinfo {author} {\bibfnamefont {David~J.}\
  \bibnamefont {Luitz}}\ and\ \bibinfo {author} {\bibfnamefont {Nicolas}\
  \bibnamefont {Laflorencie}},\ }\bibfield  {title} {\enquote {\bibinfo {title}
  {Quantum monte carlo detection of su(2) symmetry breaking in the
  participation entropies of line subsystems},}\ }\href {\doibase
  10.21468/SciPostPhys.2.2.011} {\bibfield  {journal} {\bibinfo  {journal}
  {SciPost Phys.}\ }\textbf {\bibinfo {volume} {2}},\ \bibinfo {pages} {011}
  (\bibinfo {year} {2017})}\BibitemShut {NoStop}%
\bibitem [{\citenamefont {Kitaev}\ and\ \citenamefont
  {Preskill}(2006)}]{Kitaev2006:TEE}%
  \BibitemOpen
  \bibfield  {author} {\bibinfo {author} {\bibfnamefont {Alexei}\ \bibnamefont
  {Kitaev}}\ and\ \bibinfo {author} {\bibfnamefont {John}\ \bibnamefont
  {Preskill}},\ }\bibfield  {title} {\enquote {\bibinfo {title} {Topological
  entanglement entropy},}\ }\href {\doibase 10.1103/PhysRevLett.96.110404}
  {\bibfield  {journal} {\bibinfo  {journal} {Phys. Rev. Lett.}\ }\textbf
  {\bibinfo {volume} {96}},\ \bibinfo {pages} {110404} (\bibinfo {year}
  {2006})}\BibitemShut {NoStop}%
\bibitem [{\citenamefont {Levin}\ and\ \citenamefont
  {Wen}(2006)}]{Levin2006:TEE}%
  \BibitemOpen
  \bibfield  {author} {\bibinfo {author} {\bibfnamefont {Michael}\ \bibnamefont
  {Levin}}\ and\ \bibinfo {author} {\bibfnamefont {Xiao-Gang}\ \bibnamefont
  {Wen}},\ }\bibfield  {title} {\enquote {\bibinfo {title} {Detecting
  topological order in a ground state wave function},}\ }\href {\doibase
  10.1103/PhysRevLett.96.110405} {\bibfield  {journal} {\bibinfo  {journal}
  {Phys. Rev. Lett.}\ }\textbf {\bibinfo {volume} {96}},\ \bibinfo {pages}
  {110405} (\bibinfo {year} {2006})}\BibitemShut {NoStop}%
\bibitem [{\citenamefont {Grover}\ \emph {et~al.}(2013)\citenamefont {Grover},
  \citenamefont {Zhang},\ and\ \citenamefont
  {Vishwanath}}]{Grover2013:EEPortal}%
  \BibitemOpen
  \bibfield  {author} {\bibinfo {author} {\bibfnamefont {Tarun}\ \bibnamefont
  {Grover}}, \bibinfo {author} {\bibfnamefont {Yi}~\bibnamefont {Zhang}}, \
  and\ \bibinfo {author} {\bibfnamefont {Ashvin}\ \bibnamefont {Vishwanath}},\
  }\bibfield  {title} {\enquote {\bibinfo {title} {Entanglement entropy as a
  portal to the physics of quantum spin liquids},}\ }\href {\doibase
  10.1088/1367-2630/15/2/025002} {\bibfield  {journal} {\bibinfo  {journal}
  {New Journal of Physics}\ }\textbf {\bibinfo {volume} {15}},\ \bibinfo
  {pages} {025002} (\bibinfo {year} {2013})}\BibitemShut {NoStop}%
\bibitem [{\citenamefont {Melko}\ \emph {et~al.}(2010)\citenamefont {Melko},
  \citenamefont {Kallin},\ and\ \citenamefont
  {Hastings}}]{Melko2010:MutualInfo}%
  \BibitemOpen
  \bibfield  {author} {\bibinfo {author} {\bibfnamefont {Roger~G.}\
  \bibnamefont {Melko}}, \bibinfo {author} {\bibfnamefont {Ann~B.}\
  \bibnamefont {Kallin}}, \ and\ \bibinfo {author} {\bibfnamefont {Matthew~B.}\
  \bibnamefont {Hastings}},\ }\bibfield  {title} {\enquote {\bibinfo {title}
  {Finite-size scaling of mutual information in monte carlo simulations:
  Application to the spin-$\frac{1}{2}$ $xxz$ model},}\ }\href {\doibase
  10.1103/PhysRevB.82.100409} {\bibfield  {journal} {\bibinfo  {journal} {Phys.
  Rev. B}\ }\textbf {\bibinfo {volume} {82}},\ \bibinfo {pages} {100409}
  (\bibinfo {year} {2010})}\BibitemShut {NoStop}%
\bibitem [{\citenamefont {Hastings}\ \emph {et~al.}(2010)\citenamefont
  {Hastings}, \citenamefont {Gonz\'{a}lez}, \citenamefont {Kallin},\ and\
  \citenamefont {Melko}}]{Hastings2010:MeasureREE}%
  \BibitemOpen
  \bibfield  {author} {\bibinfo {author} {\bibfnamefont {Matthew~B.}\
  \bibnamefont {Hastings}}, \bibinfo {author} {\bibfnamefont {Iv\'{a}n}\
  \bibnamefont {Gonz\'{a}lez}}, \bibinfo {author} {\bibfnamefont {Ann~B.}\
  \bibnamefont {Kallin}}, \ and\ \bibinfo {author} {\bibfnamefont {Roger~G.}\
  \bibnamefont {Melko}},\ }\bibfield  {title} {\enquote {\bibinfo {title}
  {Measuring renyi entanglement entropy in quantum monte carlo simulations},}\
  }\href {\doibase 10.1103/PhysRevLett.104.157201} {\bibfield  {journal}
  {\bibinfo  {journal} {Phys. Rev. Lett.}\ }\textbf {\bibinfo {volume} {104}},\
  \bibinfo {pages} {157201} (\bibinfo {year} {2010})}\BibitemShut {NoStop}%
\bibitem [{\citenamefont {Humeniuk}\ and\ \citenamefont
  {Roscilde}(2012)}]{Humeniuk2012:QMCREEgeneric}%
  \BibitemOpen
  \bibfield  {author} {\bibinfo {author} {\bibfnamefont {Stephan}\ \bibnamefont
  {Humeniuk}}\ and\ \bibinfo {author} {\bibfnamefont {Tommaso}\ \bibnamefont
  {Roscilde}},\ }\bibfield  {title} {\enquote {\bibinfo {title} {Quantum monte
  carlo calculation of entanglement r\'enyi entropies for generic quantum
  systems},}\ }\href {\doibase 10.1103/PhysRevB.86.235116} {\bibfield
  {journal} {\bibinfo  {journal} {Phys. Rev. B}\ }\textbf {\bibinfo {volume}
  {86}},\ \bibinfo {pages} {235116} (\bibinfo {year} {2012})}\BibitemShut
  {NoStop}%
\bibitem [{\citenamefont {Inglis}\ and\ \citenamefont
  {Melko}(2013)}]{Inglis2013:WangLandau}%
  \BibitemOpen
  \bibfield  {author} {\bibinfo {author} {\bibfnamefont {Stephen}\ \bibnamefont
  {Inglis}}\ and\ \bibinfo {author} {\bibfnamefont {Roger~G.}\ \bibnamefont
  {Melko}},\ }\bibfield  {title} {\enquote {\bibinfo {title} {Wang-landau
  method for calculating r\'enyi entropies in finite-temperature quantum monte
  carlo simulations},}\ }\href {\doibase 10.1103/PhysRevE.87.013306} {\bibfield
   {journal} {\bibinfo  {journal} {Phys. Rev. E}\ }\textbf {\bibinfo {volume}
  {87}},\ \bibinfo {pages} {013306} (\bibinfo {year} {2013})}\BibitemShut
  {NoStop}%
\bibitem [{\citenamefont {Luitz}\ \emph {et~al.}(2014)\citenamefont {Luitz},
  \citenamefont {Plat}, \citenamefont {Laflorencie},\ and\ \citenamefont
  {Alet}}]{Luitz2014:ImprovingEE}%
  \BibitemOpen
  \bibfield  {author} {\bibinfo {author} {\bibfnamefont {David~J.}\
  \bibnamefont {Luitz}}, \bibinfo {author} {\bibfnamefont {Xavier}\
  \bibnamefont {Plat}}, \bibinfo {author} {\bibfnamefont {Nicolas}\
  \bibnamefont {Laflorencie}}, \ and\ \bibinfo {author} {\bibfnamefont
  {Fabien}\ \bibnamefont {Alet}},\ }\bibfield  {title} {\enquote {\bibinfo
  {title} {Improving entanglement and thermodynamic r\'{e}nyi entropy
  measurements in quantum monte carlo},}\ }\href {\doibase
  10.1103/PhysRevB.90.125105} {\bibfield  {journal} {\bibinfo  {journal} {Phys.
  Rev. B}\ }\textbf {\bibinfo {volume} {90}},\ \bibinfo {pages} {125105}
  (\bibinfo {year} {2014})}\BibitemShut {NoStop}%
\bibitem [{\citenamefont {Jarzynski}(1997)}]{Jarzynski1997:Noneql}%
  \BibitemOpen
  \bibfield  {author} {\bibinfo {author} {\bibfnamefont {C.}~\bibnamefont
  {Jarzynski}},\ }\bibfield  {title} {\enquote {\bibinfo {title}
  {Nonequilibrium equality for free energy differences},}\ }\href {\doibase
  10.1103/PhysRevLett.78.2690} {\bibfield  {journal} {\bibinfo  {journal}
  {Phys. Rev. Lett.}\ }\textbf {\bibinfo {volume} {78}},\ \bibinfo {pages}
  {2690--2693} (\bibinfo {year} {1997})}\BibitemShut {NoStop}%
\bibitem [{\citenamefont {Crooks}(1999)}]{Crooks1999:Noneql}%
  \BibitemOpen
  \bibfield  {author} {\bibinfo {author} {\bibfnamefont {Gavin~E.}\
  \bibnamefont {Crooks}},\ }\bibfield  {title} {\enquote {\bibinfo {title}
  {Entropy production fluctuation theorem and the nonequilibrium work relation
  for free energy differences},}\ }\href {\doibase 10.1103/PhysRevE.60.2721}
  {\bibfield  {journal} {\bibinfo  {journal} {Phys. Rev. E}\ }\textbf {\bibinfo
  {volume} {60}},\ \bibinfo {pages} {2721--2726} (\bibinfo {year}
  {1999})}\BibitemShut {NoStop}%
\bibitem [{\citenamefont {Ritort}(2006)}]{Ritort2006:MolecExp}%
  \BibitemOpen
  \bibfield  {author} {\bibinfo {author} {\bibfnamefont {F}~\bibnamefont
  {Ritort}},\ }\bibfield  {title} {\enquote {\bibinfo {title} {Single-molecule
  experiments in biological physics: methods and applications},}\ }\href
  {\doibase 10.1088/0953-8984/18/32/r01} {\bibfield  {journal} {\bibinfo
  {journal} {Journal of Physics: Condensed Matter}\ }\textbf {\bibinfo {volume}
  {18}},\ \bibinfo {pages} {R531--R583} (\bibinfo {year} {2006})}\BibitemShut
  {NoStop}%
\bibitem [{\citenamefont {Alba}(2017)}]{Alba2017:Jarzynski}%
  \BibitemOpen
  \bibfield  {author} {\bibinfo {author} {\bibfnamefont {Vincenzo}\
  \bibnamefont {Alba}},\ }\bibfield  {title} {\enquote {\bibinfo {title}
  {Out-of-equilibrium protocol for r\'{e}nyi entropies via the jarzynski
  equality},}\ }\href {\doibase 10.1103/PhysRevE.95.062132} {\bibfield
  {journal} {\bibinfo  {journal} {Phys. Rev. E}\ }\textbf {\bibinfo {volume}
  {95}},\ \bibinfo {pages} {062132} (\bibinfo {year} {2017})}\BibitemShut
  {NoStop}%
\bibitem [{\citenamefont {Sandvik}(2005)}]{Sandvik2005:GSproj}%
  \BibitemOpen
  \bibfield  {author} {\bibinfo {author} {\bibfnamefont {Anders~W.}\
  \bibnamefont {Sandvik}},\ }\bibfield  {title} {\enquote {\bibinfo {title}
  {Ground state projection of quantum spin systems in the valence-bond
  basis},}\ }\href {\doibase 10.1103/PhysRevLett.95.207203} {\bibfield
  {journal} {\bibinfo  {journal} {Phys. Rev. Lett.}\ }\textbf {\bibinfo
  {volume} {95}},\ \bibinfo {pages} {207203} (\bibinfo {year}
  {2005})}\BibitemShut {NoStop}%
\bibitem [{\citenamefont {Sandvik}\ and\ \citenamefont
  {Evertz}(2010)}]{Sandvik2010:LoopVBB}%
  \BibitemOpen
  \bibfield  {author} {\bibinfo {author} {\bibfnamefont {Anders~W.}\
  \bibnamefont {Sandvik}}\ and\ \bibinfo {author} {\bibfnamefont {Hans~Gerd}\
  \bibnamefont {Evertz}},\ }\bibfield  {title} {\enquote {\bibinfo {title}
  {Loop updates for variational and projector quantum monte carlo simulations
  in the valence-bond basis},}\ }\href {\doibase 10.1103/PhysRevB.82.024407}
  {\bibfield  {journal} {\bibinfo  {journal} {Phys. Rev. B}\ }\textbf {\bibinfo
  {volume} {82}},\ \bibinfo {pages} {024407} (\bibinfo {year}
  {2010})}\BibitemShut {NoStop}%
\bibitem [{\citenamefont {Sutherland}(1975)}]{Sutherland1975:Multi}%
  \BibitemOpen
  \bibfield  {author} {\bibinfo {author} {\bibfnamefont {Bill}\ \bibnamefont
  {Sutherland}},\ }\bibfield  {title} {\enquote {\bibinfo {title} {Model for a
  multicomponent quantum system},}\ }\href {\doibase 10.1103/PhysRevB.12.3795}
  {\bibfield  {journal} {\bibinfo  {journal} {Phys. Rev. B}\ }\textbf {\bibinfo
  {volume} {12}},\ \bibinfo {pages} {3795--3805} (\bibinfo {year}
  {1975})}\BibitemShut {NoStop}%
\bibitem [{\citenamefont {Affleck}(1988)}]{Affleck1988:CriticalSUn}%
  \BibitemOpen
  \bibfield  {author} {\bibinfo {author} {\bibfnamefont {Ian}\ \bibnamefont
  {Affleck}},\ }\bibfield  {title} {\enquote {\bibinfo {title} {Critical
  behaviour of su(n) quantum chains and topological non-linear
  $\sigma$-models},}\ }\href {\doibase
  https://doi.org/10.1016/0550-3213(88)90117-4} {\bibfield  {journal} {\bibinfo
   {journal} {Nuclear Physics B}\ }\textbf {\bibinfo {volume} {305}},\ \bibinfo
  {pages} {582 -- 596} (\bibinfo {year} {1988})}\BibitemShut {NoStop}%
\bibitem [{\citenamefont {D'Emidio}\ \emph {et~al.}(2015)\citenamefont
  {D'Emidio}, \citenamefont {Block},\ and\ \citenamefont
  {Kaul}}]{Demidio2015:Renyi}%
  \BibitemOpen
  \bibfield  {author} {\bibinfo {author} {\bibfnamefont {Jonathan}\
  \bibnamefont {D'Emidio}}, \bibinfo {author} {\bibfnamefont {Matthew~S.}\
  \bibnamefont {Block}}, \ and\ \bibinfo {author} {\bibfnamefont {Ribhu~K.}\
  \bibnamefont {Kaul}},\ }\bibfield  {title} {\enquote {\bibinfo {title}
  {R\'enyi entanglement entropy of critical $\mathrm{SU}(n)$ spin chains},}\
  }\href {\doibase 10.1103/PhysRevB.92.054411} {\bibfield  {journal} {\bibinfo
  {journal} {Phys. Rev. B}\ }\textbf {\bibinfo {volume} {92}},\ \bibinfo
  {pages} {054411} (\bibinfo {year} {2015})}\BibitemShut {NoStop}%
\bibitem [{\citenamefont {Sandvik}(2010)}]{Sandvik2010:CompStud}%
  \BibitemOpen
  \bibfield  {author} {\bibinfo {author} {\bibfnamefont {Anders~W.}\
  \bibnamefont {Sandvik}},\ }\bibfield  {title} {\enquote {\bibinfo {title}
  {Computational studies of quantum spin systems},}\ }\href {\doibase
  10.1063/1.3518900} {\bibfield  {journal} {\bibinfo  {journal} {AIP Conference
  Proceedings}\ }\textbf {\bibinfo {volume} {1297}},\ \bibinfo {pages}
  {135--338} (\bibinfo {year} {2010})},\ \Eprint
  {http://arxiv.org/abs/https://aip.scitation.org/doi/pdf/10.1063/1.3518900}
  {https://aip.scitation.org/doi/pdf/10.1063/1.3518900} \BibitemShut {NoStop}%
\bibitem [{\citenamefont {Kaul}\ \emph {et~al.}(2013)\citenamefont {Kaul},
  \citenamefont {Melko},\ and\ \citenamefont {Sandvik}}]{Kaul2013:Bridging}%
  \BibitemOpen
  \bibfield  {author} {\bibinfo {author} {\bibfnamefont {Ribhu~K.}\
  \bibnamefont {Kaul}}, \bibinfo {author} {\bibfnamefont {Roger~G.}\
  \bibnamefont {Melko}}, \ and\ \bibinfo {author} {\bibfnamefont {Anders~W.}\
  \bibnamefont {Sandvik}},\ }\bibfield  {title} {\enquote {\bibinfo {title}
  {Bridging lattice-scale physics and continuum field theory with quantum monte
  carlo simulations},}\ }\href {\doibase
  10.1146/annurev-conmatphys-030212-184215} {\bibfield  {journal} {\bibinfo
  {journal} {Annual Review of Condensed Matter Physics}\ }\textbf {\bibinfo
  {volume} {4}},\ \bibinfo {pages} {179--215} (\bibinfo {year} {2013})},\
  \Eprint
  {http://arxiv.org/abs/https://doi.org/10.1146/annurev-conmatphys-030212-184215}
  {https://doi.org/10.1146/annurev-conmatphys-030212-184215} \BibitemShut
  {NoStop}%
\bibitem [{\citenamefont {Kaul}\ and\ \citenamefont
  {Sandvik}(2012)}]{Kaul2012:LargeN}%
  \BibitemOpen
  \bibfield  {author} {\bibinfo {author} {\bibfnamefont {Ribhu~K.}\
  \bibnamefont {Kaul}}\ and\ \bibinfo {author} {\bibfnamefont {Anders~W.}\
  \bibnamefont {Sandvik}},\ }\bibfield  {title} {\enquote {\bibinfo {title}
  {Lattice model for the $\mathrm{SU}(n)$ n\'eel to valence-bond solid quantum
  phase transition at large $n$},}\ }\href {\doibase
  10.1103/PhysRevLett.108.137201} {\bibfield  {journal} {\bibinfo  {journal}
  {Phys. Rev. Lett.}\ }\textbf {\bibinfo {volume} {108}},\ \bibinfo {pages}
  {137201} (\bibinfo {year} {2012})}\BibitemShut {NoStop}%
\bibitem [{\citenamefont {Mendes-Santos}\ \emph {et~al.}(2020)\citenamefont
  {Mendes-Santos}, \citenamefont {Giudici}, \citenamefont {Fazio},\ and\
  \citenamefont {Dalmonte}}]{Santos2020:VNnoWave}%
  \BibitemOpen
  \bibfield  {author} {\bibinfo {author} {\bibfnamefont {T}~\bibnamefont
  {Mendes-Santos}}, \bibinfo {author} {\bibfnamefont {G}~\bibnamefont
  {Giudici}}, \bibinfo {author} {\bibfnamefont {R}~\bibnamefont {Fazio}}, \
  and\ \bibinfo {author} {\bibfnamefont {M}~\bibnamefont {Dalmonte}},\
  }\bibfield  {title} {\enquote {\bibinfo {title} {Measuring von neumann
  entanglement entropies without wave functions},}\ }\href {\doibase
  10.1088/1367-2630/ab6875} {\bibfield  {journal} {\bibinfo  {journal} {New
  Journal of Physics}\ }\textbf {\bibinfo {volume} {22}},\ \bibinfo {pages}
  {013044} (\bibinfo {year} {2020})}\BibitemShut {NoStop}%
\end{thebibliography}%


\section{Supplemental material}

\subsection{QMC sampling and measurement}
Our task is to perform nonequilibrium simulations in the space of configurations contained in $\mathcal{Z}^{(n)}_{A}(\lambda)$ and add up all of the increments $\partial \ln g_{A}(\lambda(t),N_B(t))$ along a path between $\lambda=0$ and $\lambda=1$.  We must first describe how to stochastically sample the $\mathcal{Z}^{(n)}_{A}(\lambda)$ configurations in equilibrium.  We employ the stochastic series expansion QMC algorithm [\onlinecite{Sandvik2010:CompStud}] using the replica trick [\onlinecite{Hastings2010:MeasureREE}], where one samples $n$ independent copies of the partition function in a single simulation.  These are the configurations of $Z^{(n)}_{\o}$, which are one subset of $\mathcal{Z}^{(n)}_{A}(\lambda)$.  In order to transition to other subsets, spins in different (independent) traces need to be joined into a single trace. This can be done if the spin states in the different traces match each other.  The weight of the configuration will also change upon joining spins, since this changes the prefactor $g_{A}(\lambda,N_B) = \lambda^{N_B}(1-\lambda)^{N_{A}-N_B}$.  We take the acceptance probability of joining or splitting spins (with the requirement that the spin states must match) to be the ratio of the weights of the new and old configurations.  This gives the the acceptance probabilities
\begin{equation}
\label{eq:pjoinsplit}
P_{\text{join}}=\mathrm{min} \left\{ \frac{\lambda}{1-\lambda},1 \right \} \quad P_{\text{split}}=\mathrm{min} \left\{\frac{1-\lambda}{\lambda},1 \right\}.
\end{equation}
Each site in the $A$ subsystem is given the opportunity to join or split (depending on its current trace topology).  Once this rewiring is done, the connectivity of the configuration is fixed and one may perform a standard QMC update.

We now need to know how to measure the work increments as we move along a nonequilibrium trajectory in the space of configurations contained in $\mathcal{Z}^{(n)}_{A}(\lambda)$.  The increment between time $t_m$ and $t_{m+1}$ is given by

\begin{equation}
\begin{split}
\label{eq:dellng}
\Delta   \ln g_{A}(\lambda,N_B)=&(N_A-N_B(t_m)) \ln \left( \frac{1-\lambda(t_{m+1})}{1-\lambda(t_m)}\right) \\
&+N_B(t_m)\ln\left( \frac{\lambda(t_{m+1})}{\lambda(t_m)}  \right).
\end{split}
\end{equation}
In other words, the increment is calculated by fixing the configuration and computing the change in $\ln g_A$ as $\lambda$ is incremented.  This is in direct analogy to the way in which nonequilibrium work increments are computed in classical systems, except there one computes the change in energy of a configuration.

At this point two comments are in order.  Firstly, one can avoid numerical rounding errors incurred by summing the $\log$ increments by instead taking a product of the arguments.  Secondly, the factor of $-1/\beta$ in the definition of the work (Eq. (\ref{eq:Wd})) completely drops out of the formulation when one uses Jarzynski's equality (Eq. (\ref{eq:jarzynski})).  It is only included to give $W^{(n)}_A$ the proper units of work.  With these ideas in mind, we can succinctly write the Jarzynski estimator for the partition function ratio as follows:
\begin{equation}
\label{eq:jarzynski2}
\frac{Z_A}{Z_{\o}}=\bigg\langle\prod_m\frac{g_A(\lambda (t_{m + 1}),N_B(t_m))}{g_A(\lambda (t_m),N_B(t_m))}\bigg\rangle,
\end{equation}
where here we have suppressed the R\'{e}nyi index.

Now that we have the ingredients necessary to update and perform work measurements on our configurations, we will briefly outline the the main steps of the algorithm.  First we equilibrate the system in the $Z^{(n)}_{\o}$ ensemble of configurations (i.e. $\lambda=0$).  The equilibrated configuration is then saved to a file, and the nonequilibrium process begins.  First we measure the work increment by Eq. (\ref{eq:dellng}) (or better to multiply the $\lambda$ factors from the first time step in Eq. (\ref{eq:jarzynski2})), then the current value of $\lambda$ is incremented and each spin in the $A$ subsystem is given the opportunity to change its trace topology according the the probabilities in Eq. (\ref{eq:pjoinsplit}).  The trace topology is then held fixed and a regular QMC update is performed.  Then the cycle begins again by computing the next increment and so on. Throughout the entire nonequilibrium process all of the work increments are accumulated and saved, in practice sampling them at regular intervals.

In order to initiate a new nonequilibrium process, we first read in the saved configuration that was equilibrated at the beginning, then we re-equilibrate for some smaller number of QMC steps (always staying in the $Z^{(n)}_{\o}$ ensemble) and again save the configuration to a file.  This new starting configuration is used for the next nonequilibrium process. 

We note that if the quenches are fast and the entangling region is large, it can happen that at the end of the quench when $\lambda=1$ spins have not entirely joined and the ratio formula in Eq. (\ref{eq:jarzynski2}) gives zero.  This should be avoided by increasing the quench time, but (so as not to lose the work measurement) one can multiply by one (instead of zero) for these spins.  Although we have tried to avoided it, this does not seem to have any detectable effect on the accuracy of the entanglement entropy measurements, even for reasonably short quench times.

\subsection{Space dependent quench functions}

In order to clearly see how to treat space dependent quench functions, we write the formula for $\mathcal{Z}^{(n)}(\lambda)$ as:
\begin{equation}
\label{eq:Zlamprd}
\mathcal{Z}^{(n)}(\lambda) = \sum_{B \subseteq \mathcal{L}} \left( \prod_{x \in B} \lambda(x)  \prod_{x \in \bar{B}}[1-\lambda(x)] \right) Z^{(n)}_B.
\end{equation}
Here $B$ is now summed over all proper subsets of the entire lattice.  One can now interpolate between $Z^{(n)}_{\o}$ and $Z^{(n)}_{A}$ by using the space dependent quench function $\lambda(x,t)=\lambda(t) \chi_A(x)$, where the indicator function $\chi_A(x)=1$ if $x\in A$ (zero otherwise), and $\lambda(t)=(t-t_i)/(t_f-t_i)$.  This is the same function appearing in panel (a) of Fig. \ref{fig:qpchain1}, but formulating things in this way allows us to treat arbitrary quench functions.

Computing the work increments in this case is no more complicated.  First we define the $g$ function as
\begin{equation}
\label{eq:Ga}
g(\lambda(x),B)=\prod_{x \in B} \lambda(x)  \prod_{x \in \bar{B}}[1-\lambda(x)],
\end{equation}

and the dynamical work is

\begin{equation}
\label{eq:ginc}
W^{(n)}= -\frac{1}{\beta}\int^{t_f}_{t_i} dt \frac{d\lambda}{dt}\frac{\partial \ln g(\lambda(x,t),B(t))}{\partial \lambda}.
\end{equation}

The work increment accumulated between two adjacent time steps $t_m$ and $t_{m+1}$ is given by

\begin{equation}
\label{eq:deltaW}
\Delta \ln g =\ln\left( \frac{g(\lambda(x,t_{m+1}),B(t_m))}{g(\lambda(x,t_{m}),B(t_m))} \right), 
\end{equation}
or 
\begin{equation}
\label{eq:deltaW2}
\Delta \ln g =\ln\left( \prod_{x\in B(t_m)} \frac{\lambda(x,t_{m+1})}{\lambda(x,t_{m})}  \prod_{x\in \bar{B}(t_m)} \frac{1-\lambda(x,t_{m+1})}{1-\lambda(x,t_{m})} \right).
\end{equation}

The total work and Jarzynski estimator for the R\'enyi entanglement entropy follow naturally, and the space dependent $\lambda$ is used in the joining and splitting probabilities.

\subsection{QMC versus ED}
\begin{figure}[h]
\centerline{\includegraphics[angle=0,width=1.0\columnwidth]{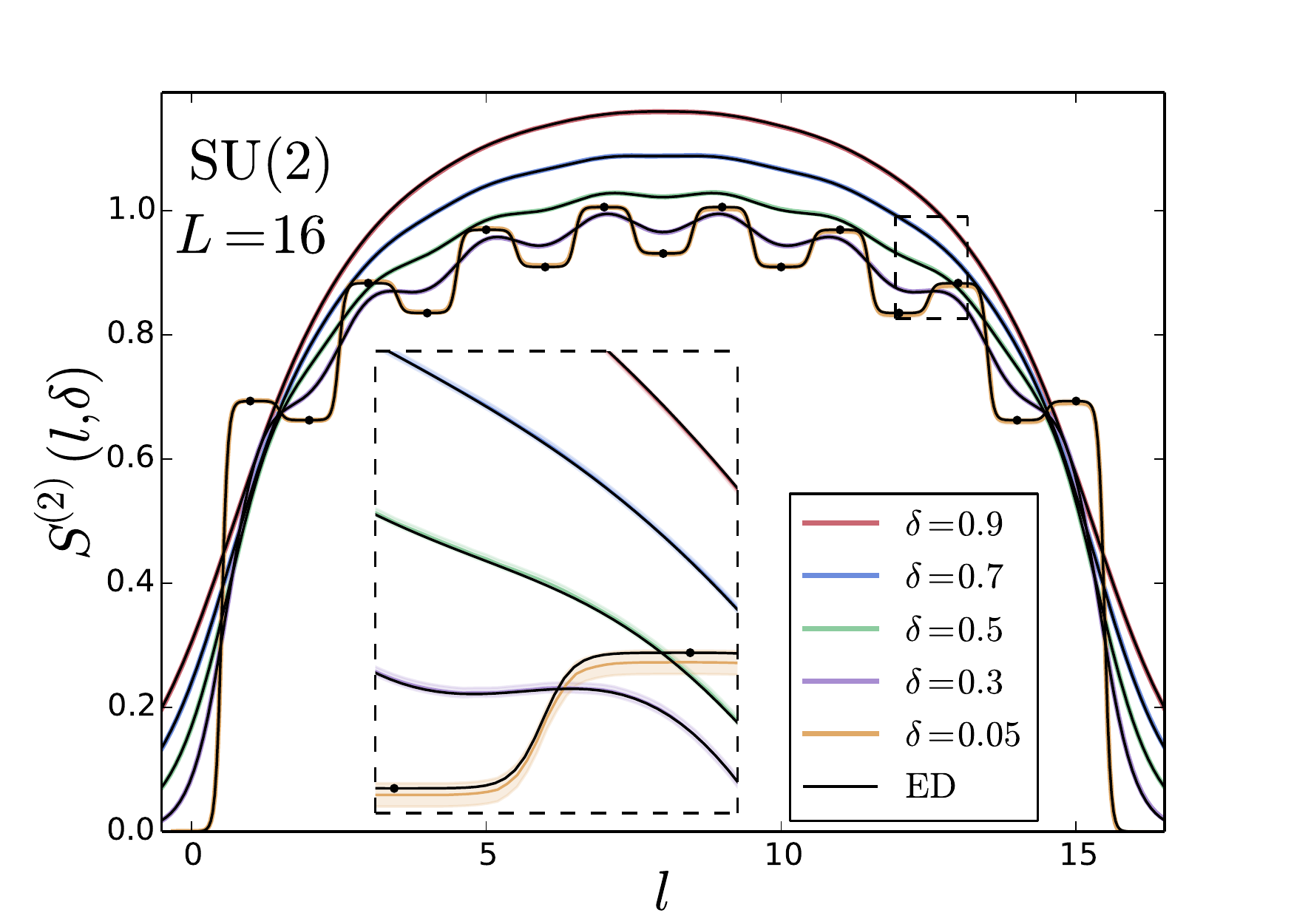}}
\caption{Here we compare our QMC method (colored curves) with results obtained from exact diagonalization (black curves and black dots) of a periodic $L=16$ site chain for $\mathrm{SU}(2)$.  We have used the quench function in Eq. (\ref{eq:lambda_l}) for different values of delta in our nonequilibrium QMC simulations and compared that with exact results with the same quench function obtained by diagonalizing the reduced density matrices for all possible bipartitions and weighting them with the appropriate factors of $\lambda(x,l,\delta)$.  We find perfect agreement between QMC and exact diagonalization, and we see that our quench function reproduces the exact second R\'enyi entanglement entropy of a block subsystem (the black dots) when $\delta$ is small.}
\label{fig:1dQMCvsED}
\end{figure}

When using any numerical method, it is always important to compare with exact results.  We have gone to great lengths to check the accuracy of our method for small systems that can be diagonalized exactly.  Fig. \ref{fig:1dQMCvsED} shows a comparison of our QMC method with exact diagonalization of an $\mathrm{SU}(2)$ $L=16$ chain.  We have used the quench function in Eq. (\ref{eq:lambda_l}) for different values of $\delta$.  The QMC data is the colored curves with shading as the error bar (this can be seen in the zoomed inset), and the black curves are exact results obtained by diagonalizing the reduced density matrices for all possible bipartitions and weighting them with the appropriate factors of $\lambda(x,l,\delta)$.  We find perfect agreement between the QMC and ED.  We also see that when $\delta$ is made small enough the exact R\'enyi entanglement entropy of a block subsystem (the black dots) is produced.

\begin{figure}[!t]
\centerline{\includegraphics[angle=0,width=1.0\columnwidth]{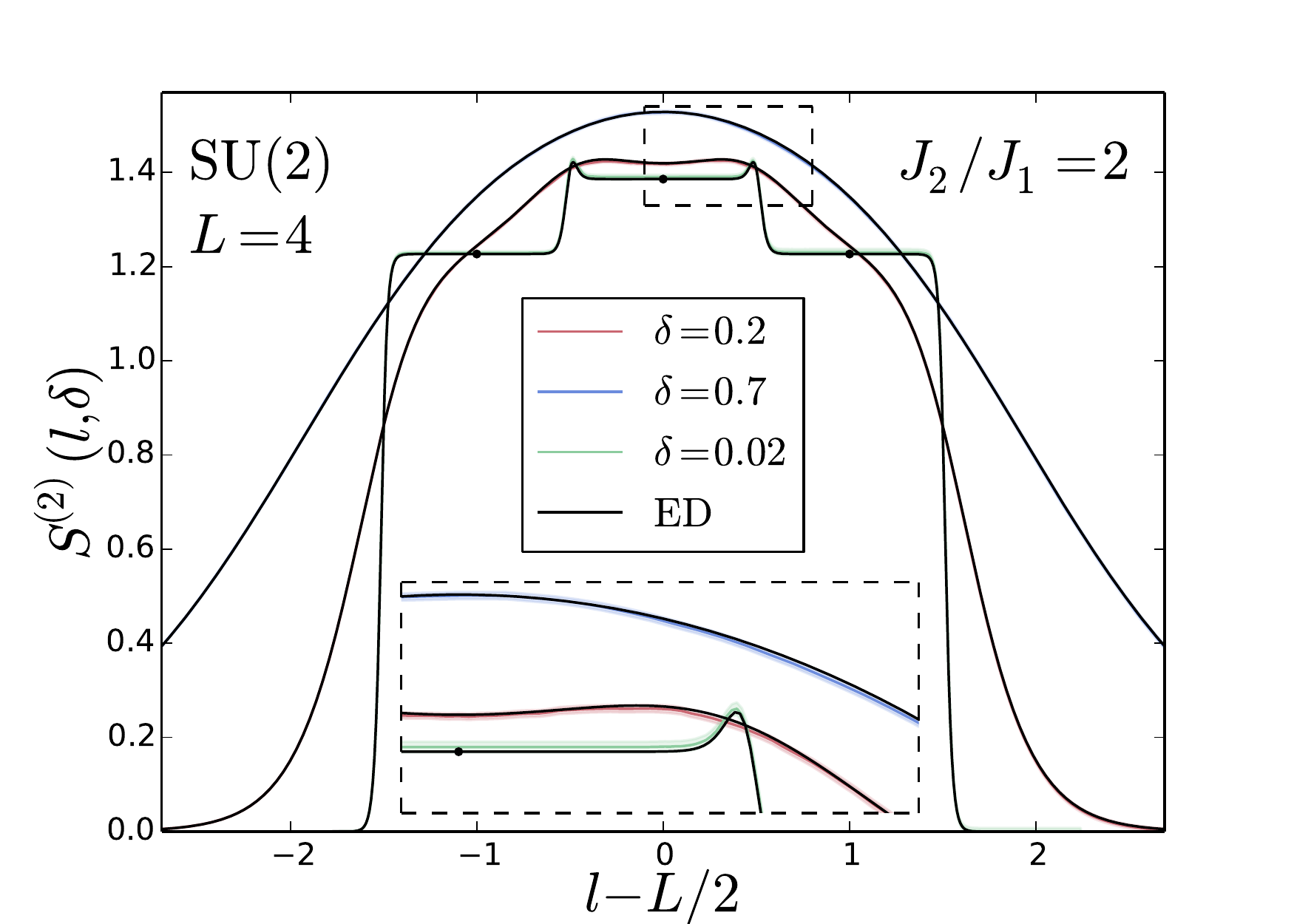}}
\caption{Here we show the same type of QMC vs ED comparison as in Fig. \ref{fig:1dQMCvsED} except applied to the two dimensional model in Eq. (\ref{eq:2DHam}) for an SU(2) L=4 system with $J_2/J_1=2$.  Here as in the main text, the same quench function is used, except in the two dimensional case it is taken to be constant in space along the y-direction.  Again when $\delta$ is small the exact second R\'enyi entanglement entropy of ribbon subsystem (black dots) is reproduced.}
\label{fig:2dQMCvsED}
\end{figure}

We also provide the same type of comparison in Fig. \ref{fig:2dQMCvsED} for our two dimensional model on an SU(2) $L=4$ square lattice with $J_2/J_1=2$ using the same quench function as in the the main text.  Again we find perfect agreement within the error bars.

\begin{figure}[!t]
\centerline{\includegraphics[angle=0,width=1.0\columnwidth]{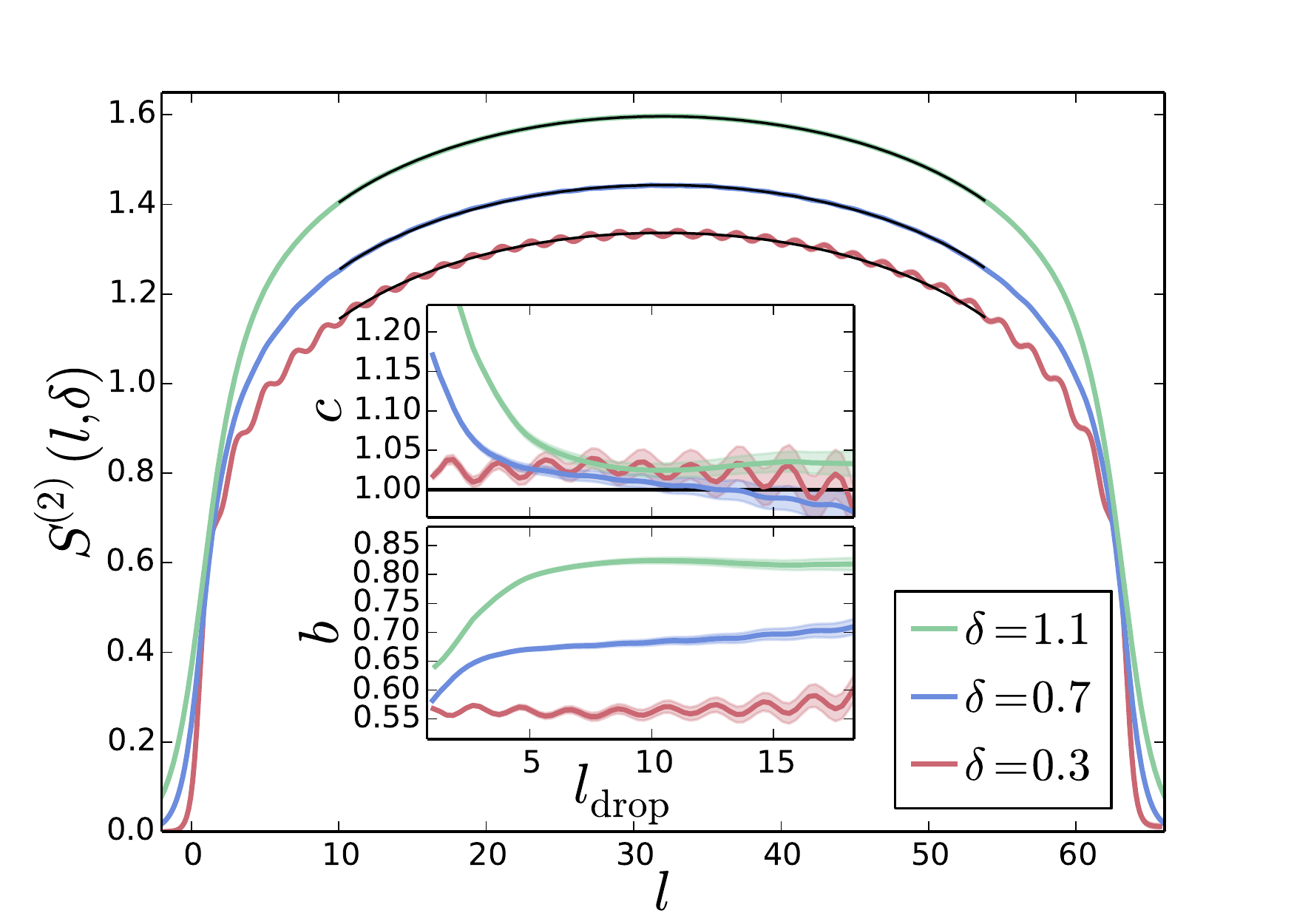}}
\caption{The second R\'enyi entanglement entropy using the quench protocol in Eq. (\ref{eq:lambda_l}) for an $\mathrm{SU}(2)$ $L=64$ periodic chain.  Here we quench the chain with three different values of $\delta$, which represents the smoothing of the subsystem's right boundary.  For $\delta=0.3$, individual sites of the chain can be resolved and one observes oscillations as a function of subsystem size.  For larger $\delta$ the contribution from the boundary is averaged over several sites and oscillations disappear.  By performing fits to the universal scaling form in Eq. (\ref{eq:LogScaling}) as a function of data dropped from the edges ($l_{\mathrm{drop}}$), we see that the estimation of the central charge ($c$) is insensitive to the presence of $\delta$ which only affects the area law constant $b$.}
\label{fig:L64su2chain}
\end{figure}

\subsection{Smooth quench fuctions and universal information}

We now wish to show that our smooth quench function preserves the universal features of the entanglement entropy.  In Fig. \ref{fig:L64su2chain} we use our newly developed nonequilibrium method  combined with the quench protocol in Eq. (\ref{eq:lambda_l}) to show that lattice scale oscillations of the entanglement entropy can be suppressed by smoothing out the subsystem boundary using the parameter $\delta$.  Crucially the central charge is insensitive to the presence of $\delta$, which only affects the nonuniversal constant (area law) contribution.

\subsection{Efficiency of Method}

\begin{figure}[t]
\centerline{\includegraphics[angle=0,width=1.0\columnwidth]{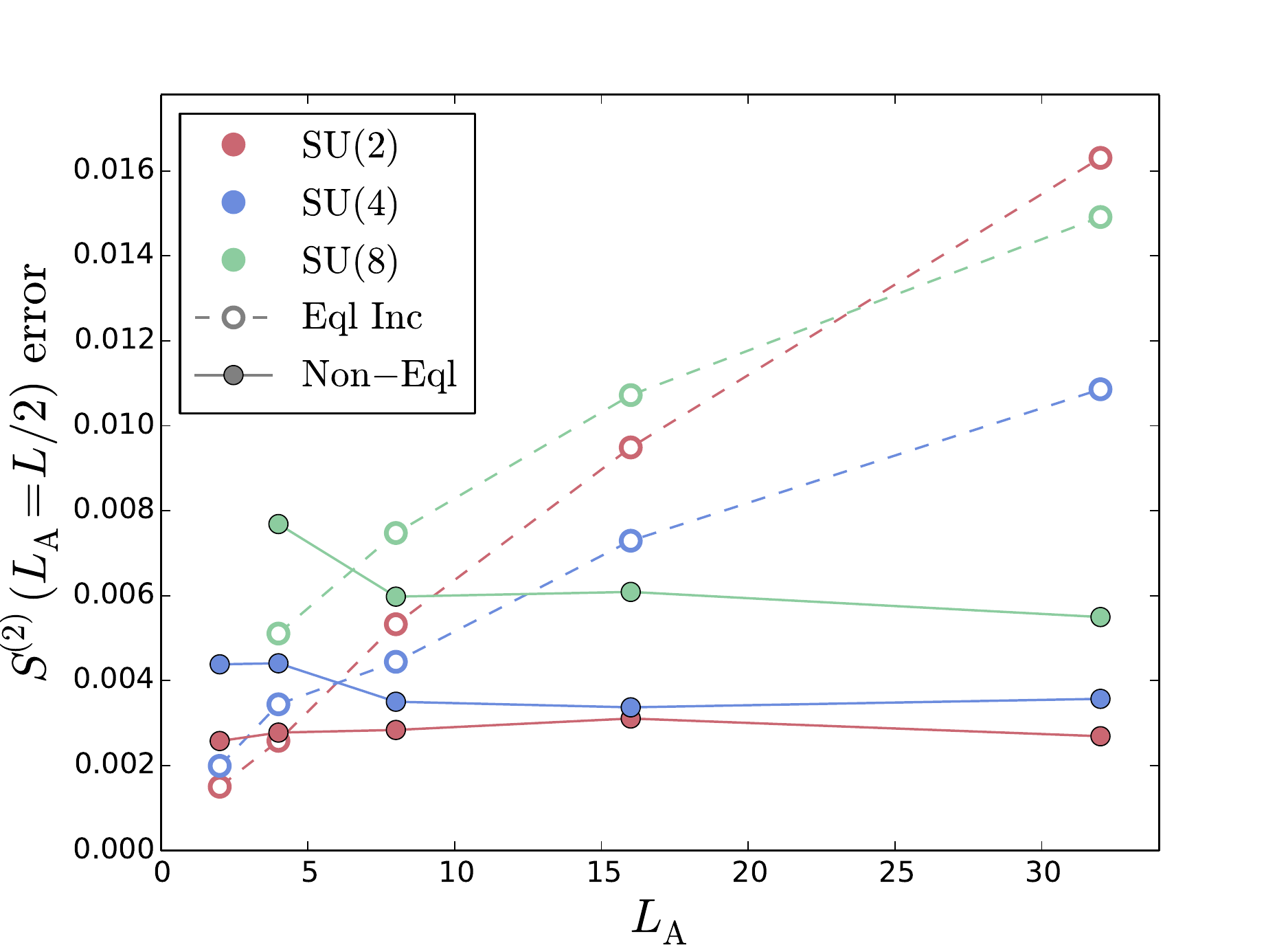}}
\caption{Here we compare our new nonequilibrium method (Non-Eql) against the extended ensemble method using the increment trick (Eql Inc).  We consider SU(N) Heisenberg chains and plot the QMC error bar (statistical error) versus the subsystem size, where we cut the chains in half fixing $L_A=L/2$.  For the extended ensemble method we build up the half chain entanglement entropy by computing the increment ratio for each site individually.  This consists of $L_A$ independent simulations.  Each simulation produced 28 x 10 binned measurements, with each measurement consisting of 10,000 sweeps.  In our nonequilibrium method, we quench the entire half chain at once using the spatially constant quench function in panel (a) of Fig. \ref{fig:qpchain1}.  For this we compute 28 x 10 independent work realizations each consisting of $L_A$ x 10,000 nonequilibrium time steps.  The total number of measurement sweeps is then identical between the two methods and we see that the nonequilibrium method remarkably maintains a constant error bar.}
\label{fig:Efficiency}
\end{figure}

\begin{figure}[t]
\centerline{\includegraphics[angle=0,width=1.0\columnwidth]{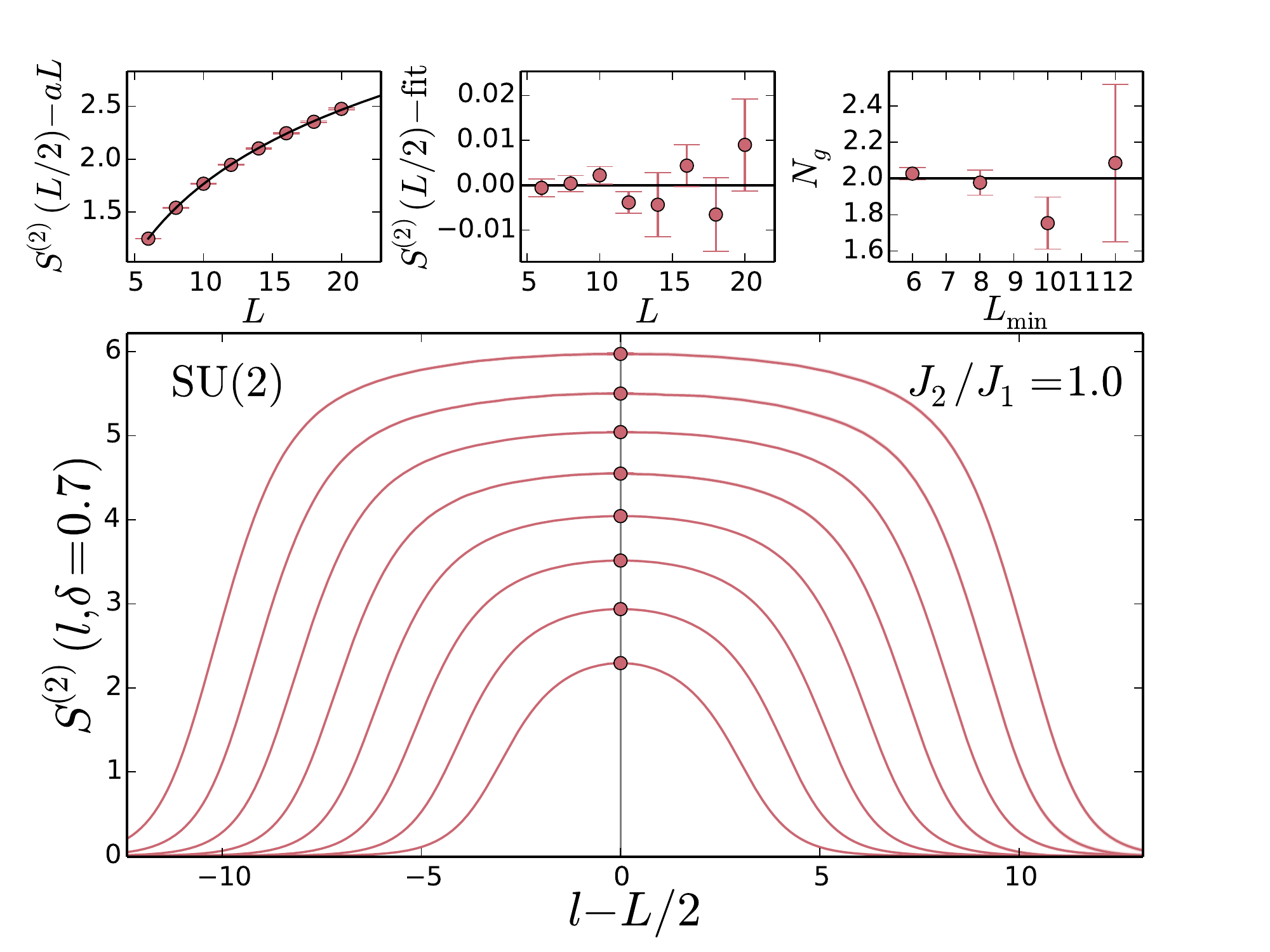}}
\caption{Here we show our zero temperature converged data for the two dimensional model for SU(2) at $J_2/J_1=1$.  The main plot shows the raw data obtained from our quench function with $\delta=0.7$, and the upper subplots (from left to right), show the center cut data with the area law piece subtracted, the center cut data minus the fit performed to Eq. (\ref{eq:goldlog}), and the extracted number of Goldstone modes ($N_g$) versus the smallest system size used in the fit ($L_{\mathrm{min}}$).}
\label{fig:su2Binf}
\end{figure}

\begin{figure}[t]
\centerline{\includegraphics[angle=0,width=1.0\columnwidth]{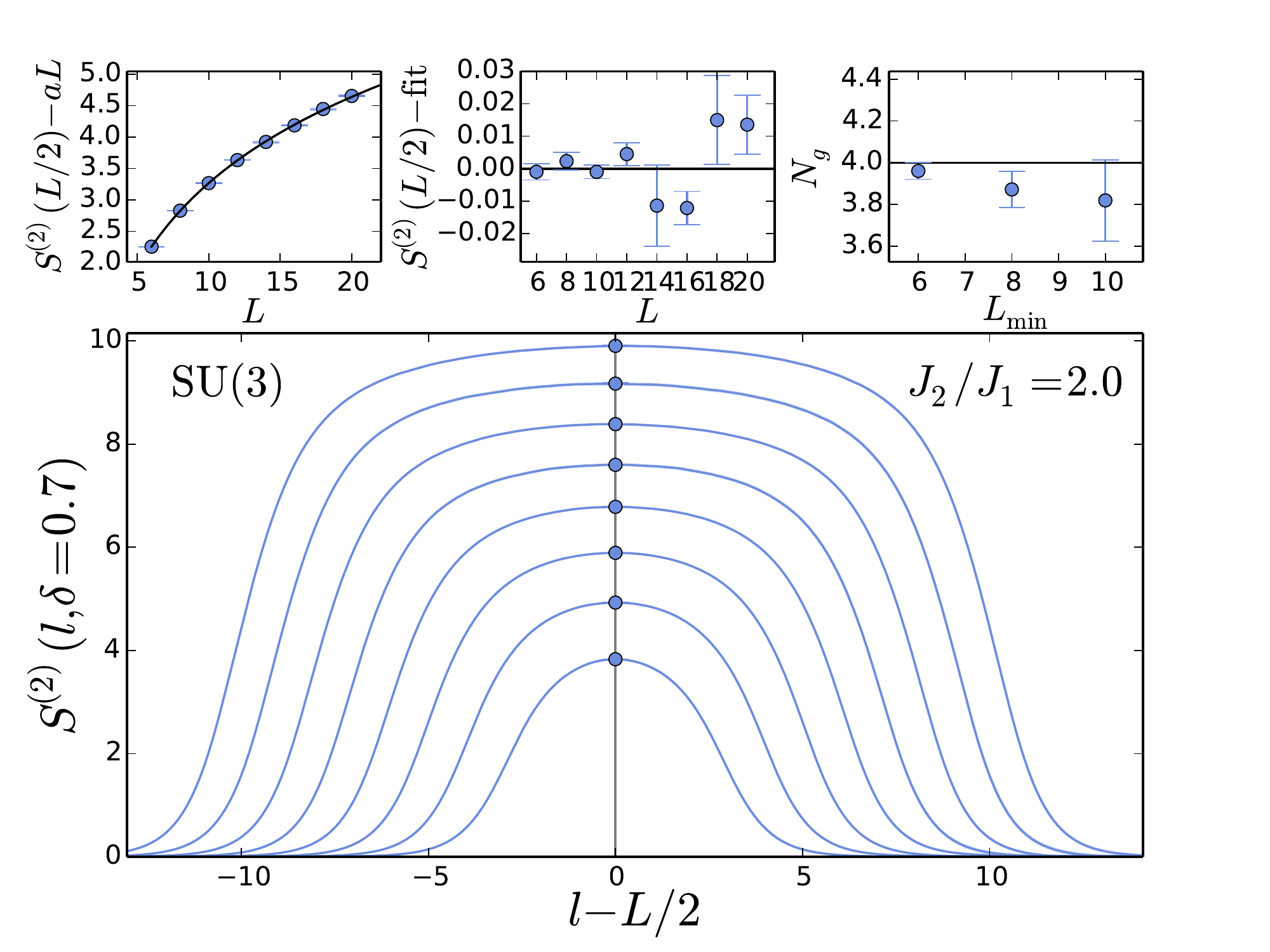}}
\caption{This figure is similar to Fig. \ref{fig:su2Binf} except for SU(3) at $J_2/J_1=2$.}
\label{fig:su3Binf}
\end{figure}

\begin{figure}[t]
\centerline{\includegraphics[angle=0,width=1.0\columnwidth]{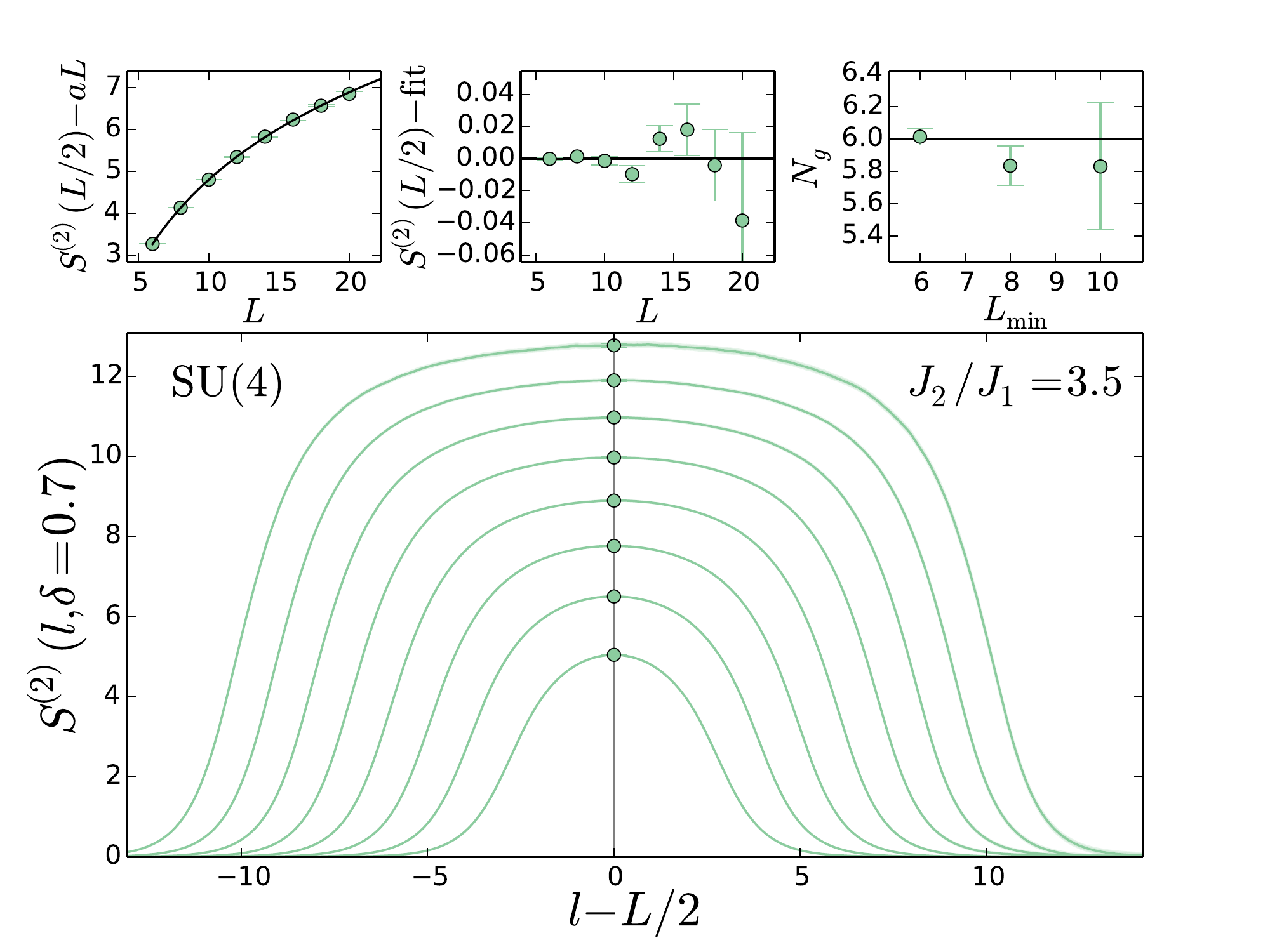}}
\caption{This figure is similar to Fig. \ref{fig:su2Binf} except for SU(4) at $J_2/J_1=3.5$.)}
\label{fig:su4Binf}
\end{figure}

It is clear from the results presented in the main text that this method is very efficient.  But here we would like to give a direct comparison of our nonequilibrium quenches with the extended ensemble method [\onlinecite{Humeniuk2012:QMCREEgeneric}] combined with the increment trick [\onlinecite{Hastings2010:MeasureREE}].  In Fig. \ref{fig:Efficiency} we show the statistical error (QMC error bar) as a function of subsystem size $L_A=L/2$ for $\mathrm{SU}(N)$ chains.  Here we compute the second R\'enyi entanglement entropy of half of the chain in two ways: (1) by directly quenching the entire half chain using our nonequilibrium method and (2) by using the extended ensemble method with increments of one site at a time.  For the extended ensemble method, we compute each increment in a separate (equilibrium) simulation with 28 x 10 binned measurements each consisting of 10,000 measurement sweeps.  For our nonequilibrium method we quench the entire half chain at once with 28 x 10 work realizations each consisting of $L_A$ $\mathrm{x}$ 10,000  nonequilibrium time steps.  As such, the total number of measurement sweeps used for each method is identical.  Very remarkably, in the nonequilibrium case the error bar stays flat as long as the number of nonequilibrium time steps is increased in proportion to the number of sites quenched.   Significant computational resources are also saved on equilibration, since the independent equilibrium increments need to be separately equilibrated.

\subsection{$T=0$ converged data for 2D model}

Here we would like to provide a more detailed view of the data presented in the main paper for the two dimensional SU($N$) model.  In order to obtain $T=0$ converged data, we have had to set $\beta=L^2$ with $J_1=1$.  These extremely low temperatures prohibit us from simulating systems much larger than $L=20$.  In Fig. \ref{fig:su2Binf} we again show the raw data for $\mathrm{SU}(2)$ appearing in the main text, this time with subfigures that show (from left to right) the $\log$ contribution on linear axes, the QMC data with the fit subtracted, and the fitted value of $N_g$ as a function of the smallest system size used.  We show the same type of plot for $\mathrm{SU}(3)$ and $\mathrm{SU}(4)$ in Fig. \ref{fig:su3Binf} and Fig. \ref{fig:su4Binf}, respectively.

\subsection{Finite-$T$ data for 2D model}

\begin{figure}[h]
\centerline{\includegraphics[angle=0,width=1.0\columnwidth]{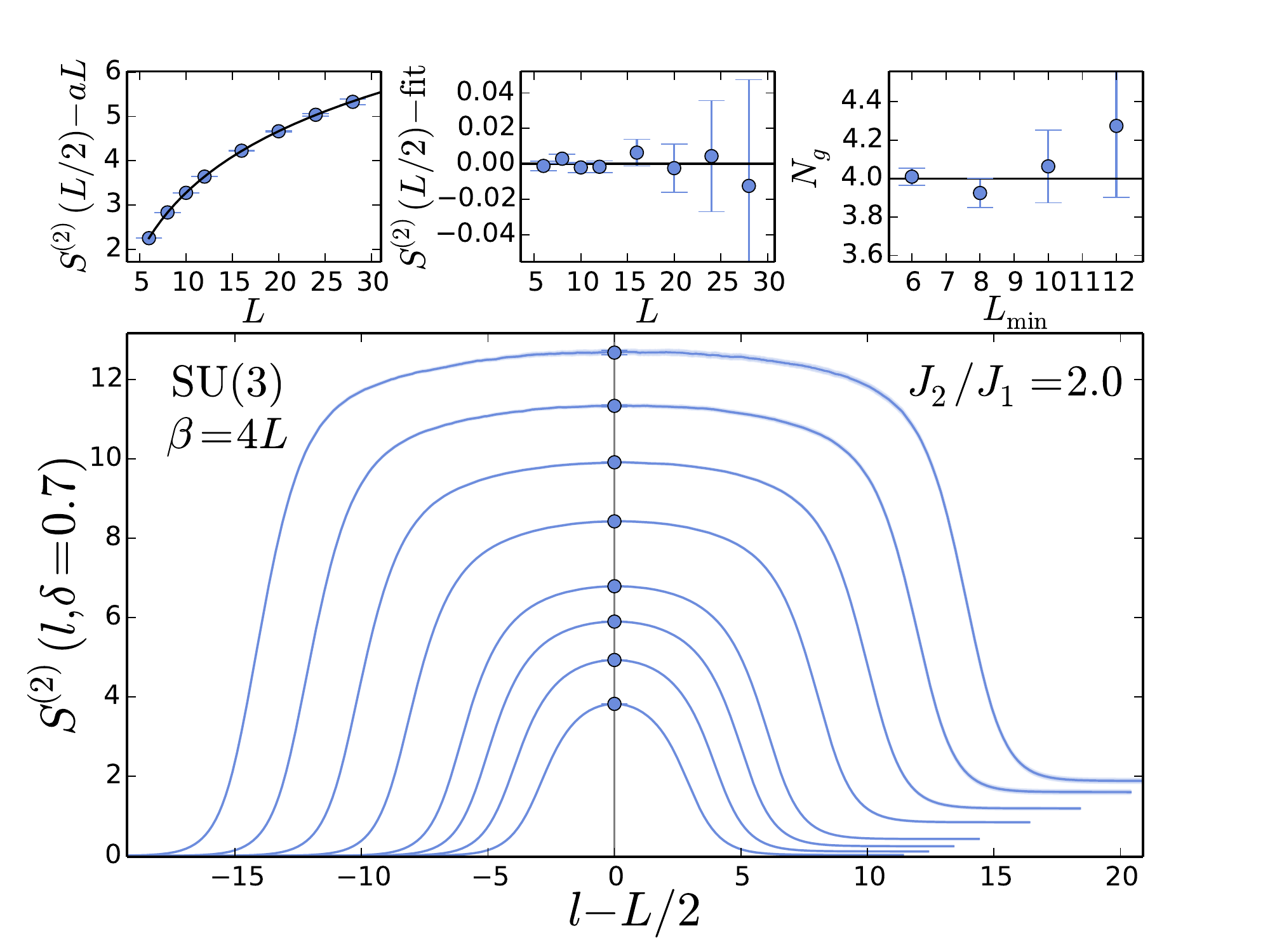}}
\caption{This figure is similar to Fig. \ref{fig:su2Binf} except computed at finite temperature, taking $\beta=4L$ with $J_1=1$ and $J_2/J_1=2$ for SU(3).  Here finite temperature affects are clearly visible since the density matrix is no longer pure.  We are able to reach slightly larger system sizes in this case, and our numerical fits look to be even higher quality than in the $T=0$ case.  We conclude from this that $N_g$ can be reliably extracted even from finite temperature data.}
\label{fig:su3B4L}
\end{figure}

\begin{figure}[h]
\centerline{\includegraphics[angle=0,width=1.0\columnwidth]{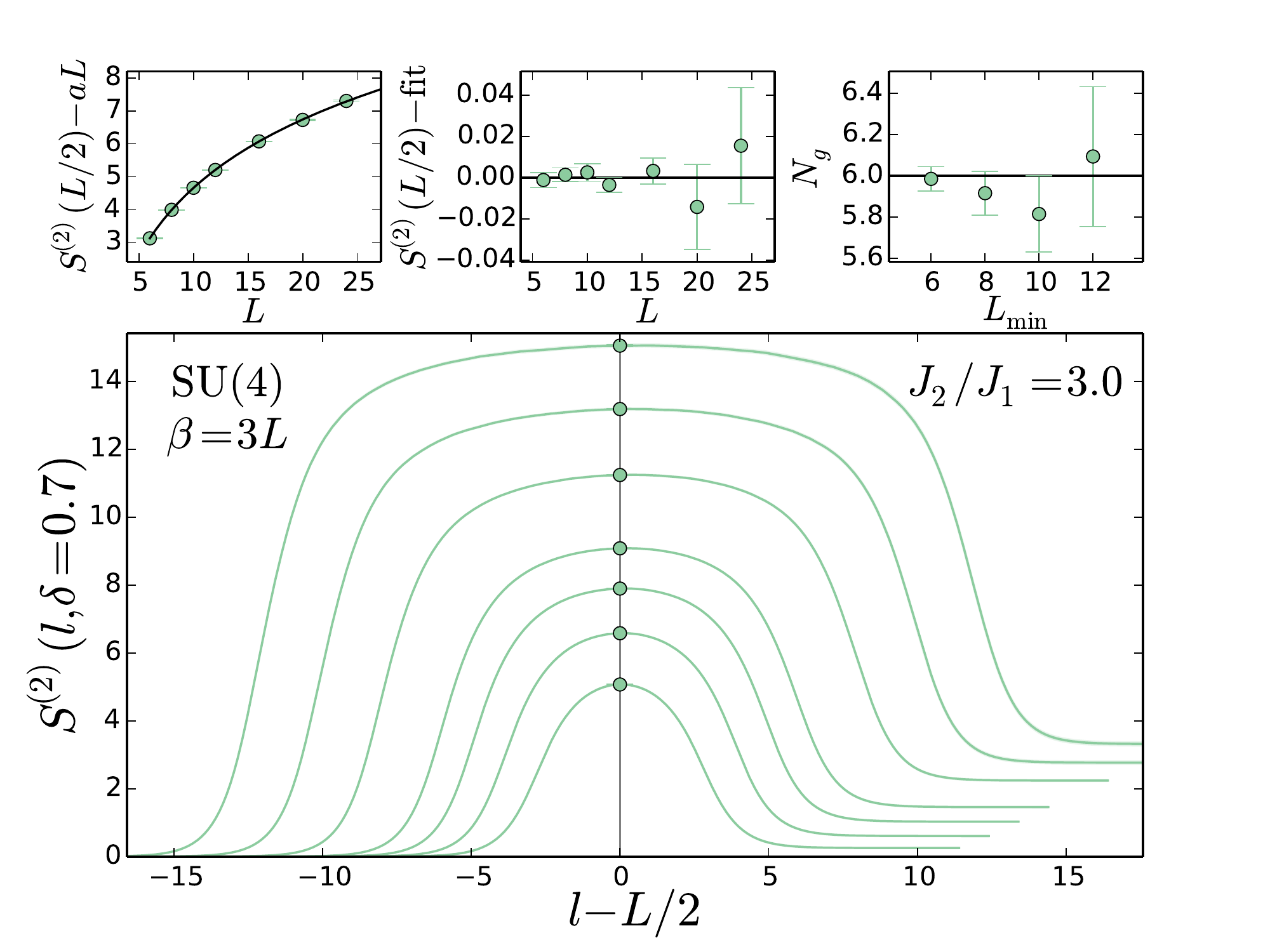}}
\caption{This figure is the same as Fig. \ref{fig:su3B4L} except for SU(4) with $\beta=3L$ and $J_2/J_1=3$.  Again we find a very high quality fit even at finite temperature.}
\label{fig:su4B3L}
\end{figure}

We have also collected data at finite temperatures, this time only scaling $\beta$ proportional to $L$ and not $L^2$.  Remarkably, the finite temperature effects do not influence the extraction of $N_g$ as we show in Fig. \ref{fig:su3B4L} for $\mathrm{SU}(3)$ and Fig. \ref{fig:su4B3L} for $\mathrm{SU}(4)$.  These results at finite temperature arguably produce even more high quality fits than the data converged at $T=0$. 

\begin{figure}[h]
\centerline{\includegraphics[angle=0,width=1.0\columnwidth]{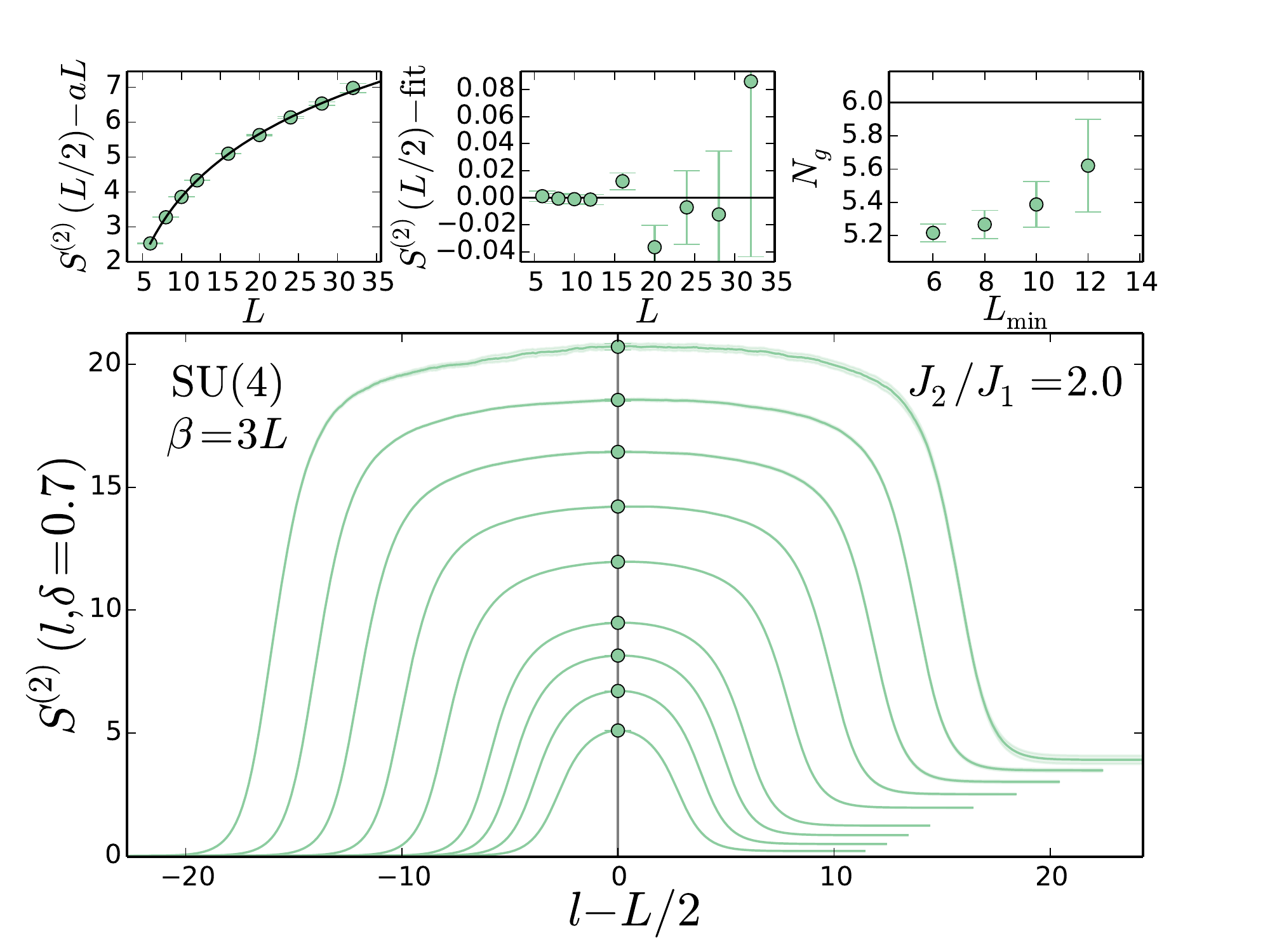}}
\caption{This figure is the same as Fig. \ref{fig:su4B3L} except we now set $J_2/J_1=2$ to illustrate the presence of finite size effects when $J_2$ is too small.  We can see a systematic drift toward $N_g=6$ as smaller system sizes are excluded from the fit.}
\label{fig:su4B3Lv2}
\end{figure}

Finally we include data for SU(4) at a lower value of $J_2/J_1$ to illustrate the finite size effects in this case.  Fig. \ref{fig:su4B3Lv2} shows finite temperature data collected with $\beta=3L$ and $J_2/J_1=2$.  We can see a systematic dependence of $N_g$ on $L_{\mathrm{min}}$ in our numerical fits, with a drift toward the true value $N_g=6$.

\begin{figure}[h]
\centerline{\includegraphics[angle=0,width=1.0\columnwidth]{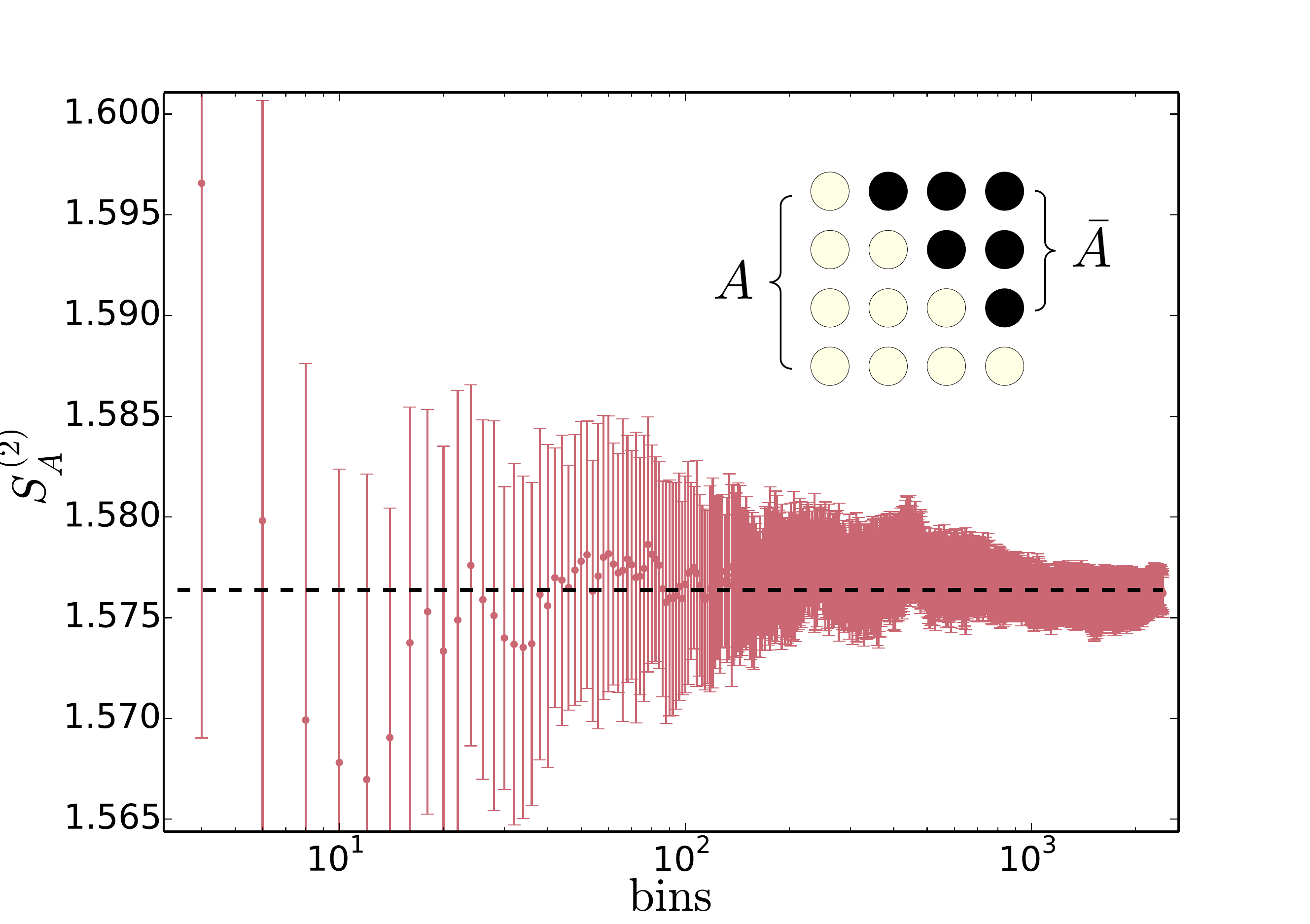}}
\caption{The second R\'{e}nyi entanglement entropy for the Heisenberg antiferromagnet on a 4 x 4 square lattice, choosing the $A$ subsystem as pictured.  We find perfect agreement with the exact result (dashed line) as a function of the number of bins used for the averaging.}
\label{fig:T0projED}
\end{figure}

\subsection{Projector QMC simulations}

We now briefly outline how the nonequilibrium R\'{e}nyi entanglement entropy measurements can be extended to projector QMC simulations in the valence bond basis [\onlinecite{Sandvik2005:GSproj}, \onlinecite{Sandvik2010:LoopVBB}].  Here instead of using Monte Carlo techniques to sample the quantum partition function at finite temperature, one instead chooses a trial wave function that is projected into the ground state by acting with powers of the Hamiltonian.  The main advantage (aside from some additional technical simplifications) is that very good trial wavefunctions can be used, so that the number of operators needed to reach the ground state is far fewer compared to the stochastic series expansion at low temperatures.

In the projector method we use the fact that the unnormalized ground state $|\psi\rangle$ can be written as
\begin{equation}
\label{eq:psi}
|\psi\rangle = \lim_{m\to\infty}H^m|\psi_{\text{trial}}\rangle.
\end{equation}
One can then write the density matrix (dividing by the norm) and use it to construct the R\'{e}nyi entanglement entropy for a region $A$.  Just as with the replica partitions functions, the R\'{e}nyi entanglement entropy will take the form of a log of a ratio.  In this case, however, it will be a ratio of wave function overlaps.  For the second R\'{e}nyi entropy of a region $A$, we can write it as follows:
\begin{equation}
\label{eq:R2swap}
 S^{(2)}_A = -\ln\left( \frac{  \langle \psi | \text{Swap}_A |\psi\rangle \langle \psi | \text{Swap}_A |\psi\rangle }{ \langle \psi  |\psi\rangle \langle \psi  |\psi\rangle} \right),
\end{equation}
where $ \text{Swap}_A$ means that the spin degrees of freedom have been exchanged in the region $A$ between the two overlaps in the numerator.

\begin{figure}[h]
\centerline{\includegraphics[angle=0,width=1.0\columnwidth]{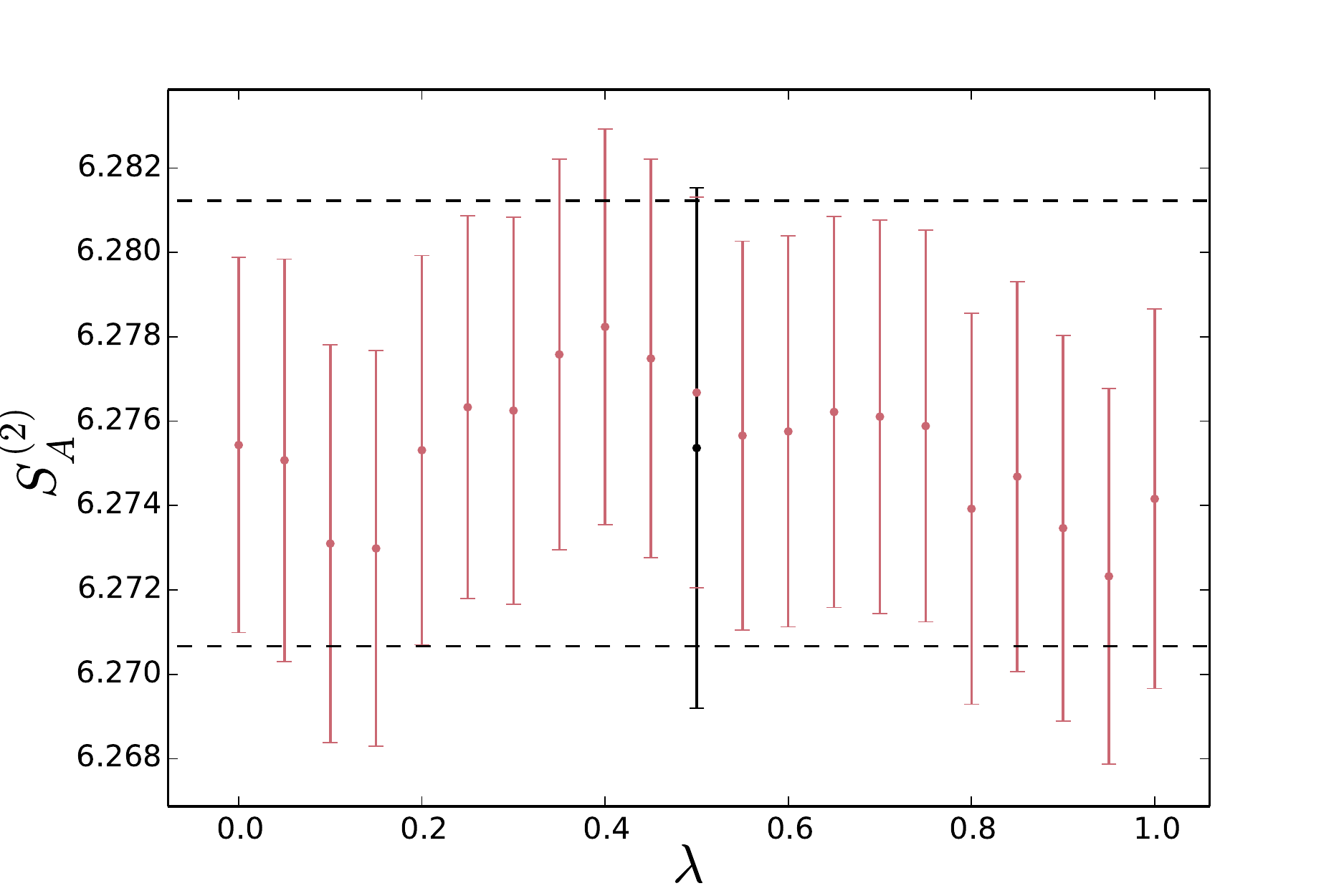}}
\caption{The half-system second R\'{e}nyi entanglement entropy for the Heisenberg antiferromagnet on a 24 x 24 square lattice.  We have combined the forward and reverse work measurements to obtain the half-system entropy at each point along the nonequilibrium trajectory (here 21 points have been sampled).  All data points are statistically consistent, and the final estimate (black data point) is taken as the average.  The error on the final estimate is conservatively taken as the average error of all the points plus the standard deviation of the points away from the mean.  The dashed window is the same measurement reported in [\onlinecite{Kallin2011:AnomaliesEEHeisen}].}
\label{fig:T0projCOMP}
\end{figure}

\begin{figure}[h]
\centerline{\includegraphics[angle=0,width=1.0\columnwidth]{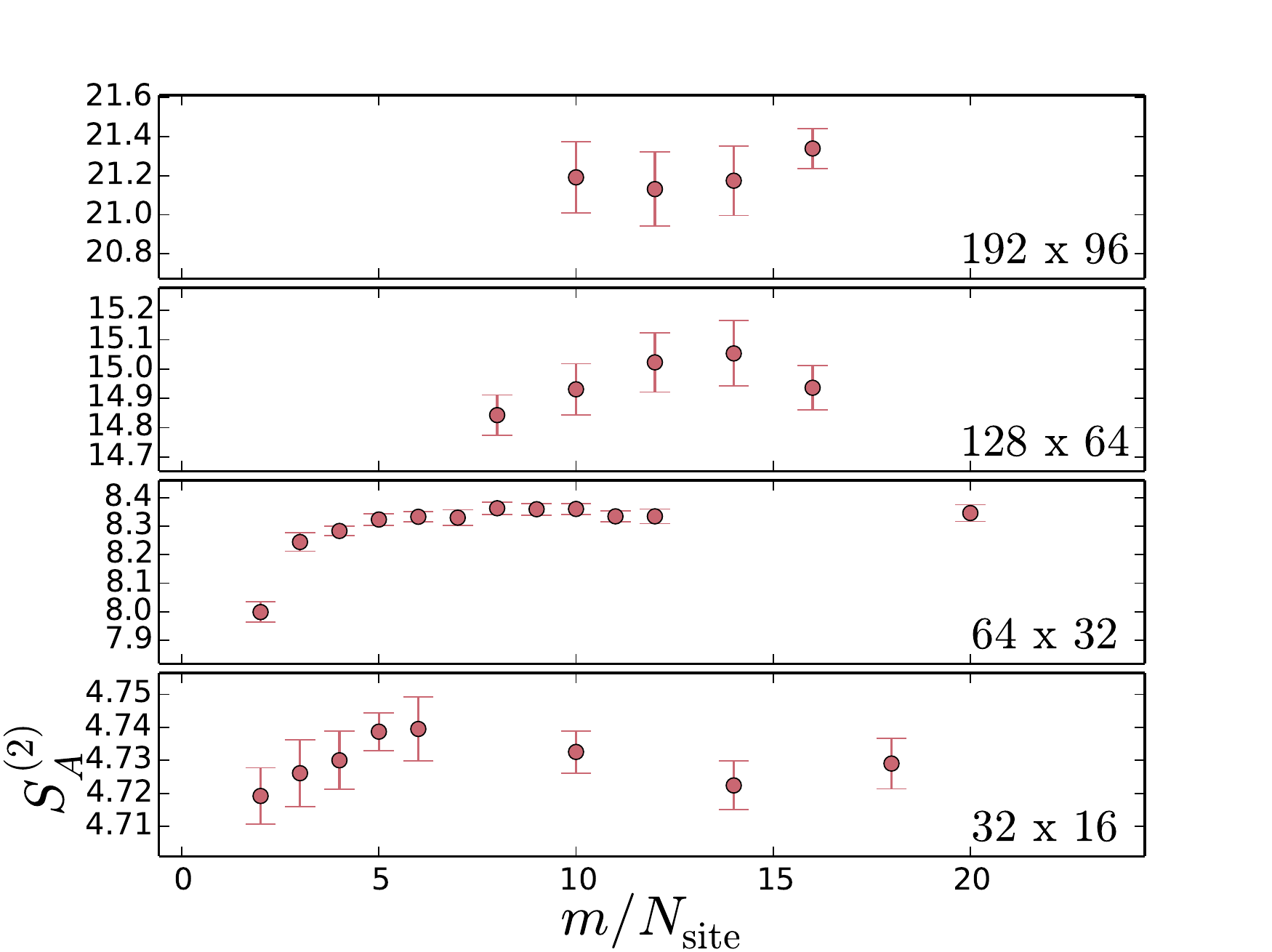}}
\caption{The half-system second R\'{e}nyi entanglement entropy for the Heisenberg antiferromagnet on different size lattices as a function of the projection power $m$.  We see that for convergence, $m$ should be taken proportional to the number of sites and that the constant of proportionality shows a slight increase as a function of system size.  We find that the entanglement entropy monotonically approaches the ground state value.}
\label{fig:T0projvsM}
\end{figure}

We now find ourselves in exactly the same situation as with the finite temperature simulations, we just need to vary an external field from 0 to 1 in such a way that it takes us from configurations contained in the denominator of Eq. (\ref{eq:R2swap}) to configurations contained in the numerator.  To achieve this we can couple the external field $\lambda$ to the single site swap operator on each site of the $A$ subsystem.  We use the same prefactor $g_A$ as before, and the measurement process is identical.  To be clear, during the simulation single site swap operators are inserted or removed throughout the $A$ subsystem with the probabilities in Eq. (\ref{eq:pjoinsplit}) whenever the spin degrees of freedom match in the center of both wave function overlaps.

For the 2D Heisenberg antiferromagnet we choose to work in the combined basis [\onlinecite{Sandvik2010:LoopVBB}] of $S^z$ spin values and valence bond (singlet) coverings of the square lattice.  Working with valence bonds allows for the construction of good trial states for the Heisenberg ground state and using the $S^z$ spin values enables efficient loop updates of the configurations.  We take our trial states to be valence bond coverings with a $1/r^3$ potential [\onlinecite{Sandvik2010:LoopVBB}].

Firstly in order to verify our $T=0$ projector code, we make a comparison with the second R\'{e}nyi entanglement entropy obtained by exact diagonalization on a 4 x 4 (periodic) lattice, which is shown in Fig. \ref{fig:T0projED}.

In order to ensure that our measurements are consistent (especially on very large system sizes), we have performed separate nonequilibrium work calculations in both the forward and reverse directions.  We find perfect agreement in all cases.  The forward and reverse measurements can be combined so that the final result is obtained at each point along the trajectory ($\lambda$ between 0 and 1).  The average of all of these points can be used as the final average and a conservative estimate of the error bar can be taken as the average error bar of all the points plus the standard deviation of the points from the mean.  This is illustrated in Fig. \ref{fig:T0projCOMP} for a 24 x 24 system that has been cut in half, which we compare with the same measurement reported in [\onlinecite{Kallin2011:AnomaliesEEHeisen}].

Finally, in Fig. \ref{fig:T0projvsM} we show the convergence of our large scale entanglement data as a function of the projection power $m$.  Here $m$ needs to be scaled with the number of sites, and the proportionality constant increases with the system size.  We find that the entanglement monotonically approaches the value in the ground state.

\end{document}